\documentclass[12pt]{iopart}

\usepackage{here}
\usepackage{graphicx}
\usepackage{epsfig}
\usepackage{color}

\newcommand{\lp}{\left(}
\newcommand{\rp}{\right)}

\newcommand{\be}{\begin{displaymath}}
\newcommand{\ee}{\end{displaymath}}
\newcommand{\bn}{\begin{equation}}
\newcommand{\en}{\end{equation}}

\newcommand{\lang}{\left\langle}
\newcommand{\rang}{\right\rangle}

\newcommand{\gyro}{{\sc gyro }}
\begin{document}

\title[Possible mechanism responsible for observed impurity outward
flow under radio frequency heating]{Possible mechanism responsible for observed
  impurity outward flow under radio frequency heating}

\author{S. Moradi, T. F\"ul\"op, A. Moll\'en, I. Pusztai}

\address{Department of Applied Physics, Nuclear Engineering, Chalmers
  University of Technology and Euratom-VR Association, G\"oteborg,
  Sweden}
\begin{abstract}
  The effect of poloidal asymmetry of impurities on impurity transport
  driven by electrostatic turbulence in tokamak plasmas is
  analyzed. It is found that in the presence of in-out asymmetric
  impurity populations the zero-flux impurity density gradient (the
  so-called peaking factor) is significantly reduced. A sign change in
  the impurity flux may occur if the asymmetry is sufficiently
  large. This may be a contributing reason for the observed outward
  convection of impurities in the presence of radio frequency
  heating. The effect of in-out asymmetry is most significant in
  regions with low temperature gradients. In the trapped electron mode
  dominated case also an up-down asymmetry can lead to a sign change
  in the peaking factor from positive to negative. The effect of ion
  parallel compressibility on the peaking factor is significant, and
  leads to positive peaking factors in regions with high temperature
  gradients, even in the presence of in-out asymmetry.
\end{abstract}
\maketitle

\section{Introduction}
Accumulation of impurities in the core of fusion devices would have
detrimental effect on fusion reactivity due to increased radiation
losses and plasma dilution. Significant effort has therefore been
spent during recent years to identify plasma conditions in which
accumulation can be avoided. One of the most promising methods for
obtaining flat or hollow impurity density profiles is to use radio
frequency heating. This has been shown to work well in various
experiments
\cite{valisa,duxppcf,duxnm,neu,puiatti03,puiatti06,angionipop07} but
the physical mechanism by which the change of the direction of the
impurity convective velocity occurs has not yet been clearly
identified in spite of the various efforts that have been made
\cite{ap,angionippcf07,nordman}.  In particular the reason for the
flat impurity density profiles in ion cyclotron resonance heated
(ICRH) discharges in JET has been debated for many years
\cite{valisa,puiatti03,puiatti06}.  In a recent paper \cite{poloidal}
it was shown that a poloidal asymmetry could lead to a significant
reduction of the impurity zero-flux density gradient (also called the
peaking factor), and even a sign change in the impurity flux, if the
asymmetry is sufficiently large. This suggests a possible explanation
for the avoidance of accumulation of high-$Z$ impurities with ICRH.

In this paper a numerical investigation of the effect of poloidal
asymmetry on impurity transport is presented. In particular, the
dependence of the impurity density peaking factor on charge number,
inverse ion- and electron temperature scale lengths and inverse electron
density scale length is analyzed. Also, the role of parallel ion
compressibility is studied. The results are benchmarked to \gyro
\cite{gyro} in the poloidally symmetric case.

It is found that inboard accumulation gives rise to negative peaking
factors (outward impurity convection and hollow impurity profile),
both in ion temperature gradient (ITG) and trapped electron (TE) mode
driven turbulence. The sign and magnitude of the peaking factor will
be shown to be sensitive not only to the asymmetry strength but also
to the temperature gradient, and it is more pronounced in the low
temperature gradient region characteristic for the plasma core. The
peaking factor is expected to be negative for moderate to high
impurity charge numbers (above $Z \simeq 15$) for JET-like
experimental parameters \cite{valisa}. In TE mode dominated cases
up-down asymmetry can also give rise to negative peaking factors if
the electron temperature gradient is sufficiently large.  As noted in
previous work \cite{ap}, in the poloidally symmetric case, that
parallel compressibility has a significant effect on the peaking
factor. As we will see in this paper this effect is even more
pronounced for asymmetric impurity densities, and will lead to
negative peaking factors for lower asymmetry strengths than without
taking into account parallel compressibility.

The remainder of the paper is organized as follows. In
Sec.~\ref{sec:poloidal} we describe the mechanism behind the impurity
poloidal asymmetry that arises in the presence of ICRH. In
Sec.~\ref{sec:flux} the model for calculating the quasilinear impurity
flux and the peaking factor in the presence of poloidal asymmetry is
presented. In Sec.~\ref{sec:pf} the parametric dependences of the
peaking factor are analyzed by presenting scans over relevant
parameters such as charge number, and temperature and density scale
lengths. Also, the importance of the impurity parallel compressibility
is demonstrated. Finally, the results are discussed and summarized in
Sec.~\ref{sec:conclusions}.

\section{Poloidal asymmetry}
\label{sec:poloidal}
Poloidal impurity asymmetries in tokamaks can arise due to various
reasons: e.g. difference in impurity source location, toroidal
rotation or neoclassical effects.  There is a wealth of experimental
evidence for poloidal asymmetries
\cite{ingessonppcf,marr,condrea,ricenf97,romanelli,sunn}. In the
plasma core in-out asymmetries can arise due to the presence of radio
frequency (RF) heating (henceforth ``in-out'' and ``out-in''
asymmetries will refer to the situations when the maximum of the
poloidally varying impurity density is at the inboard and outboard
sides of the plasma, respectively).  A detailed physical explanation
of why ICRH favors inboard accumulation, together with the description
of the experimental results, is given in Ref.~\cite{ingessonppcf}. The
underlying principle is that the asymmetry is a result of the increase
of the hydrogen-minority density on the outboard side. These particles
tend to be trapped on the outside of the torus and the turning points
of their orbits drift towards the resonance layer due to the
heating. The poloidal asymmetry in the hydrogen-minority density gives
rise to an electric field that pushes the other ion species to the
inboard side. In the case of highly-charged impurities, this effect is
amplified by their higher charge $Z$.

The RF-induced accumulation of minority ions on the outboard side
leads to a corresponding impurity accumulation on the inboard side by
the following mechanism.  If the plasma consists of electrons, bulk
ions, impurity ions and RF-heated minority ions then it can be
expected that all species except the minority ions are Boltzmann
distributed (the dynamics of the minority ions is strongly affected by
the heating). If a particle species $a$ is Boltzmann distributed, the
poloidal variation of the density is $\tilde{n}_a/n_{a0}\simeq -e_a 
\phi/T_a$, where the tilde denotes the variation on the flux surface,
$e_a$ is the charge and $T_a$ is the temperature of the
species. Quasineutrality requires that
$$
n_{e0}\left(1+\frac{e\phi}{T_e}\right)-n_{D0}\left(1-\frac{e \phi}{T_i}\right)-Zn_{Z0}\left(1-\frac{Ze\phi}{T_z}\right)=n_{H0}+\tilde{n}_H,
$$ where the subscript zero indicates the density where the potential
$\phi$ vanishes.  If $\phi$ is normalized so that
$n_{D0}+n_{H0}+Zn_{Z0}-n_{e0}=0$, the poloidal variation of the
impurity density becomes
\begin{equation}
\frac{\tilde{n}_z}{n_{Z0}}=-\frac{Ze\phi}{T_z}=-\frac{Z\tilde{n}_H/n_{D0}}{1+(T_i/T_e)+(n_{Z0}Z^2/n_{D0})}.
\label{nzpert}
\end{equation}
Since the poloidal variation in the impurity density has the opposite
sign to that of the minority ions, the accumulation of the latter on
the outboard side gives rise to an electric field that pushes the
other ion species to the inboard side.  Simulations of the hydrogen
ion distribution function in the presence of RF heating with the Monte
Carlo code FIDO \cite{fido} show that a considerable out-in asymmetry
in the hydrogen ion density can be expected, which is sufficient to
account for the observed in-out asymmetry in the impurity density.

In the tokamak edge, where the plasma is sufficiently collisional,
also steep radial pressure or temperature gradients can give rise to
an in-out asymmetry.  These effects have been observed in e.g. Alcator
C-Mod \cite{marr}, and it has been shown that the observations are in
qualitative agreement with neoclassical theory
\cite{helander,fh1,fh2,landreman}. The sign and magnitude of these asymmetries
depend sensitively and nonlinearly on magnetic geometry, fraction of
impurities in the plasma and rotation. Neoclassical theory also
predicts an up-down asymmetry, which is caused by the ion-impurity
friction.

\section{Impurity flux}
\label{sec:flux}
Since impurity transport is usually dominated by drift-wave
turbulence, in this work we focus on the effect of the impurity
poloidal asymmetry on impurity transport driven by
microinstabilities. We assume that the processes that cause the
asymmetry are not affected significantly by the fact that the
cross-field transport is dominated by fluctuations.  For simplicity we
consider only the collisionless, electrostatic case.  The quasilinear
impurity particle flux is given by
\begin{equation} 
\Gamma_z=-\frac{k_\theta}{B}\mathrm{Im}\left[\hat{n}_z\phi^\ast\right]
,\label{ge}
\end{equation}
where $\mathrm{Im}[\cdot]$ denotes imaginary part, $k_\theta$ is the poloidal
wave-number, $\hat{n}_z$ is the perturbed impurity density, $\phi^\ast$ is
the complex conjugate of perturbed electrostatic potential $\phi$.

The perturbed impurity density response in an axisymmetric, large
aspect ratio torus with circular magnetic surfaces can be obtained
from the linearized gyrokinetic equation \cite{romanellibriguglio}.
In the absence of collisions and in the limit $v_\parallel/q
R(\omega-\omega_{Dz})\ll 1$, the gyrokinetic equation can be solved
iteratively to find the non-adiabatic part of the perturbed impurity
distribution
$$
g_z=\left(1-\frac{v_\parallel}{q R(\omega-{\omega}_{Dz})}\frac{\partial}{\partial
  \theta}\frac{v_\parallel}{qR(\omega-{\omega}_{Dz})} \frac{\partial}{\partial
  \theta}\right)\frac{Ze f_{z0}}{T_z}\frac{\omega-\omega_{\ast
    z}^T}{\omega-{\omega}_{Dz}} J_0(z_z)\phi,
$$ where $\theta$ is the extended poloidal angle, $f_{a0}=n_a/\lp\sqrt{\pi}v_{Ta}
\rp^{3}\exp(-x_a^2) $ is the equilibrium Maxwellian distribution
function, $x_a=v/v_{Ta}$ is the velocity normalized to the thermal
speed $v_{Ta}=(2T_a/m_a)^{1/2}$, $n_a$ and $m_a$  are the
density  and mass  of species $a$, $\omega_{\ast
  a}=-k_\theta T_a/e_aB L_{na}$ is the diamagnetic frequency, $
\omega_{\ast a}^T=\omega_{\ast
  a}\left[1+\left(x_a^2-\frac{3}{2}\right)L_{na}/L_{Ta}\right] $,
$L_{na}=-[\partial (\ln{n_a})/\partial r]^{-1}$, $L_{Ta}=-[\partial
  (\ln{T_a})/\partial r]^{-1}$, are the density and temperature scale
lengths, $ \omega_{Da}=-k_\theta
\left(v_{\perp}^2/2+v_{\parallel}^2\right) D(\theta)/\omega_{ca} R$ is
the magnetic drift frequency, $D(\theta)= \left(\cos{\theta}+ s \theta
\sin{\theta}\right)$, $\omega_{ca}=e_a B/m_a$ is the cyclotron
frequency, $B$ is the equilibrium magnetic field, $q$ is the safety
factor, $s=(r/q)(dq/dr)$ is the magnetic shear, $r$ and $R$ are the
minor and major radii, $J_0$ is the Bessel function of the first kind
and $z_a=k_\perp v_{\perp}/\omega_{ca}$.  Including the Boltzmann part
of the distribution, the perturbed ion density response becomes
\cite{romanellibriguglio,pusztai}
\begin{eqnarray}\hspace{-2.5cm}\frac{\hat{n}_z}{n_z}= -\frac{Ze \phi}{T_z} + \int d^3 v
  \frac{Zef_{z0}J_0(z_z)}{n_z T_z}  \left[1-\frac{v_\parallel}{qR(\omega-{\omega}_{Dz})}\frac{\partial}{\partial
  \theta}\frac{v_\parallel}{qR(\omega-{\omega}_{Dz})} \frac{\partial}{\partial
  \theta}\right] \frac{ \omega-\omega_{\ast z}^T}{\omega-{\omega}_{Dz}}J_0(z_z) \phi,\nonumber\\
\label{nz}
\end{eqnarray}
where parallel compressibility is represented by the term containing
the two $\theta$-derivatives.

\begin{figure}[htbp]
\begin{center}
  \includegraphics[width=0.45\textwidth]{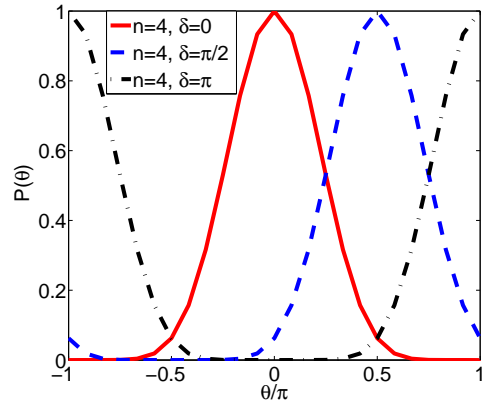}
  \caption{$\mathcal{P}(\theta)$ as a function of normalized
    $\theta/\pi$, for different values of the peaking angle
    $\delta$. The asymmetry strength is $n=4$. Solid line (red)
    represents out-in, dashed line (blue) is up-down and
    dash-dotted line (black) is in-out asymmetry.}
\label{pas}
\end{center}
\end{figure}

The zero-flux impurity density gradient (the peaking factor) can be
obtained from $\lang \Gamma_z\rang =0$ with the perturbed impurity
density taken from (\ref{nz}). Here
$\lang\cdots\rang=(1/2\pi)\int_{-\pi}^{\pi}(\cdots)d\theta $. 
If the impurity density varies over the flux surface, in general we
can write $n_z=n_{z0} \mathcal P(\theta)$, where $\mathcal
P(\theta)=1$ if the impurities are evenly distributed over the flux
surface. A poloidal asymmetry can be modelled by $\mathcal
P(\theta)=\left[\cos^2{\left(\frac{\theta-\delta}{2}\right)}\right]^n$,
where $\delta$ is the angular position where the impurity density has
its maximum.  This asymmetry function was used also in previous work
\cite{poloidal,condrea}. Figure \ref{pas} shows the asymmetry function
$\mathcal{P}(\theta)$ as a function of normalized $\theta/\pi$, for
$n=4$ and different values of $\delta$.

The impurity flux can be calculated numerically, by solving the
velocity-space integrals in the perturbed impurity density (\ref{nz})
in the expression for the impurity flux (\ref{ge}), without the
constant energy resonance approximation or the assumption on the
smallness of the finite Larmor radius parameter.  Then the impurity
peaking factor $a/L_{nz}^0$ can be obtained by setting the impurity
flux to zero, or $ \langle\mathrm{Im}[\hat{n}_z\phi^\ast]
\rangle=0$. Here, the fluctuating density is given by Eq.~(\ref{nz}),
$a$ is the outermost minor radius, and $a/L_{nz}^0$ is calculated
as \begin{eqnarray}
  a/L_{nz}^0=\frac{\langle\mathrm{Im}[S_{T}(\theta)\phi^\ast]\rangle
  }{\langle\mathrm{Im}[S_{n}(\theta)\phi^\ast]\rangle },\end{eqnarray}
where
\begin{eqnarray}
  \hspace{-2.5cm} S_{n}(\theta)= \int
  d^3x_ze^{-x_z^2}J_0[z_z(\theta)]\left(1-\delta_p\mathcal M_\theta
  \right)\frac{\bar{\omega}_{*z}
    \mathcal{P}(\theta)\phi(\theta)J_0[z_z(\theta)]}{\bar{\omega}-\bar{\omega}_{Dz}(\theta)},\label{pfn}
\end{eqnarray}
and
\begin{eqnarray}
  \hspace{-2.5cm} S_{T}(\theta)= \int
  d^3x_ze^{-x_z^2} J_0[z_z(\theta)]\left(1-\delta_p\mathcal
  M_\theta\right)\frac{[\bar{\omega} - \bar{\omega}_{*z}(x_z^2 -
    3/2)a/L_{Tz}]
    \mathcal{P}(\theta)\phi(\theta)J_0[z_z(\theta)]}{\bar{\omega}-\bar{\omega}_{Dz}(\theta)}
  ,\label{pfT}
\end{eqnarray}
with the operator representing parallel compressibility
\begin{equation}
\mathcal{M}_\theta=\frac{x_{z\parallel}}{R(\theta)(\bar{\omega}-\bar{\omega}_{Dz}(\theta))}\frac{\partial}{\partial \theta}\frac{x_{z\parallel}}{R(\theta)(\bar{\omega}-\bar{\omega}_{Dz}(\theta))}\frac{\partial}{\partial \theta},
\label{mtheta}
\end{equation}
where $R(\theta)=R_0(1+\epsilon \cos\theta)$ with $\epsilon=r/R_0$,
$\delta_p=2a^2 m_{i}/(q^2m_{z}\tau_z)$, $\tau_z=T_e/T_z$,  $b =
(\rho_s/Z) \sqrt{2 A_{Z}/\tau_zA_{i}}$, $\rho_s$ is the ion sound Larmor radius,
$\bar{\omega}_{*z}=-k_{\theta}\rho_{s} /Z\tau_z$,
$\bar{\omega}_{Dz}=-2
k_{\theta}\rho_{s}(a/R)(x^2_{z\bot}/2+x^{2}_{z\parallel})D(\theta)/Z\tau_z$,
$z_z(\theta)=x_{z\bot}bk_{\bot}(\theta)$, and
$k_{\bot}=k_{\theta}\sqrt{1+s^2 \theta^2}$. All frequencies marked
with bars are in $c_{s}/a$ units, where $c_s$ is the ion sound speed, and the main ion and impurity
temperature gradients are assumed to be equal $a/L_{Tz}=a/L_{Ti}$.

\section{Parametric dependences of the peaking factor}
\label{sec:pf}
In the calculations presented in this section we have used the
following local profile and magnetic geometry parameters: $r/a=0.3$,
$R/a=3$, $k_\theta \rho_s=0.3$, $q=1.7$, $a/L_{ne}=1.3$,
$a/L_{ni}=0.86$, $T_i/T_e=0.85$, $a/L_{Te}=0.2$, $a/L_{Ti}=0.3$,
$s=0.22$ and $\rho_s/a= 0.0035$.  These parameters have been taken
from a typical JET-shot with ICRH \cite{valisa}. The impurities are
assumed to be present in trace quantities, in the sense that
$Zn_z/n_e\ll 1$ ($n_z/n_e=10^{-3}$ is used in the simulations). This
is the baseline case in our study, and these parameters will be used
unless otherwise stated. The perturbed electrostatic potential and
eigenvalues have been obtained by linear electrostatic gyrokinetic
calculations with \gyro$\!$.
\subsection{Charge number dependence}
\begin{figure}[htbp]
\begin{center}
  \includegraphics[width=0.45\textwidth]{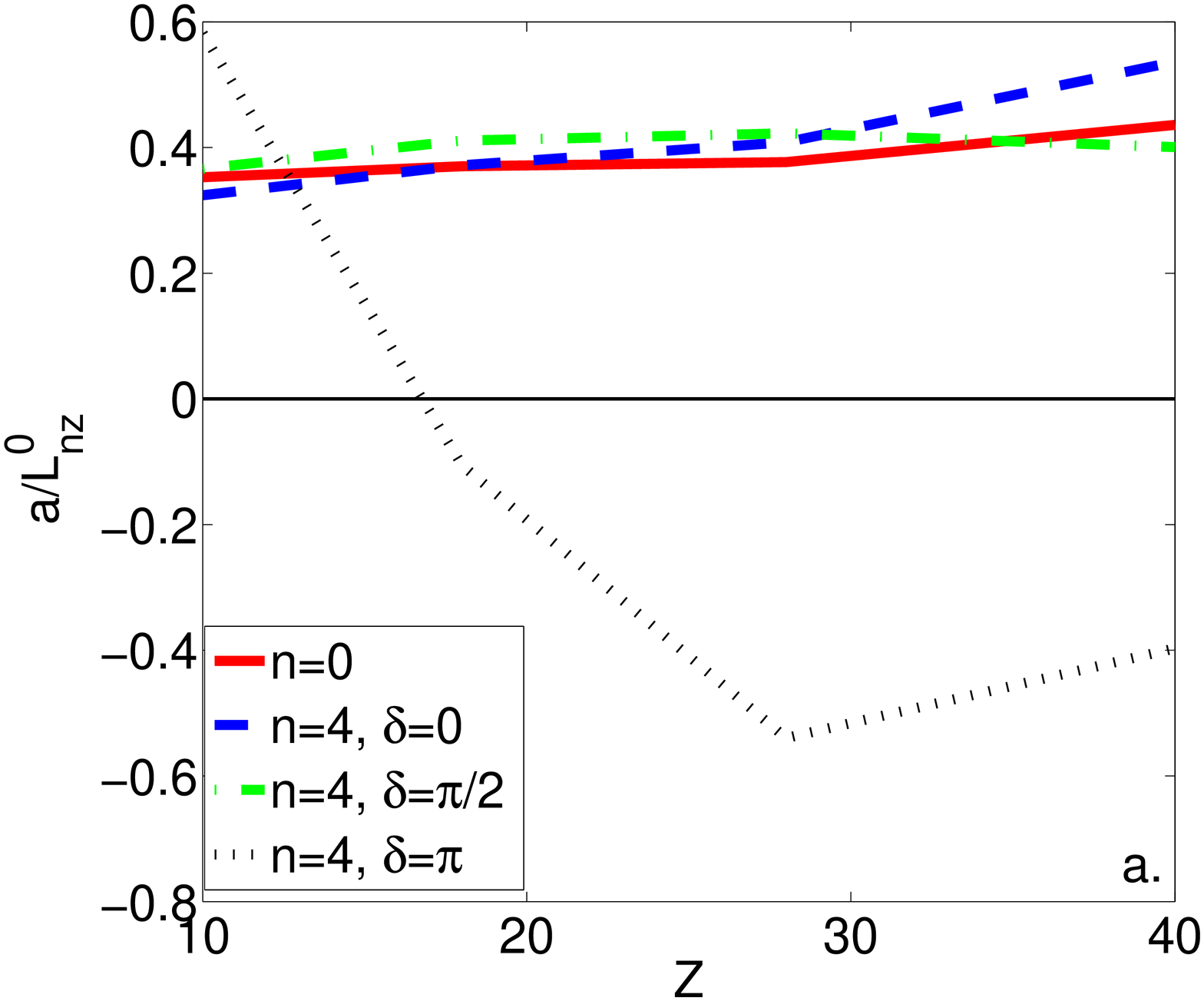}
  \includegraphics[width=0.45\textwidth]{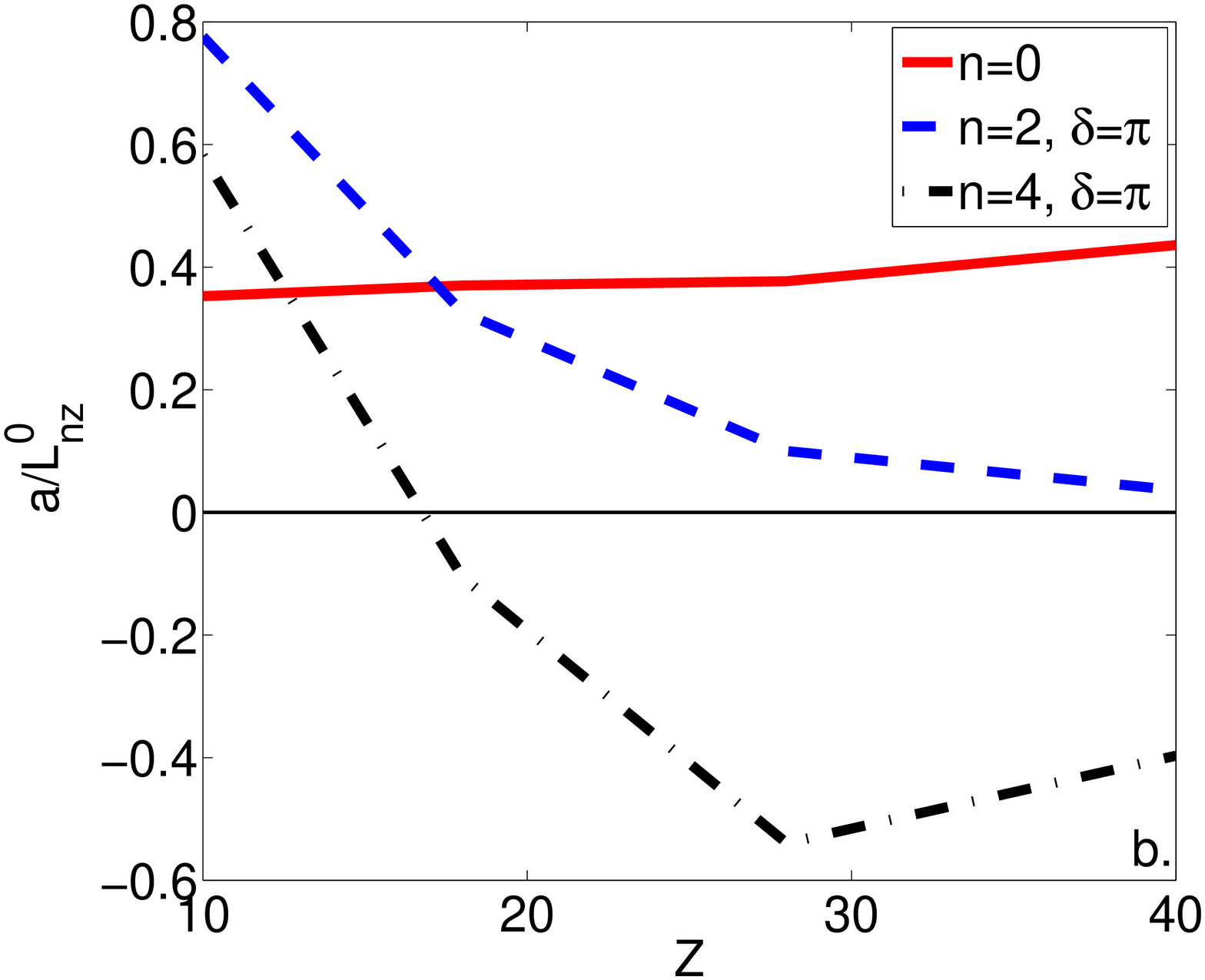}\caption{
    Peaking factor as a function of charge number for different values
    of peaking angle $\delta$ (a) and asymmetry strength $n$ (b). The
    rest of the parameters are taken from the baseline case, for which
    the eigenvalue is $\omega=(0.001+i\;0.068) c_s/a$. In both figures
    the solid line represents the case of poloidally symmetric
    impurity distribution. (a): $n=4$ -- out-in asymmetry (dashed),
    up-down asymmetry (dash-dotted), in-out asymmetry (dotted). (b):
    in-out asymmetry -- $n=2$ (dashed), $n=4$ (dash-dotted).  }
\label{zscan}
\end{center}
\end{figure}

The peaking factor as a function of charge
number for various impurity poloidal asymmetries is shown in Fig.~\ref{zscan}.
In the poloidally symmetric case (solid line in Fig.~\ref{zscan}a),
the peaking factor is not sensitive to the charge number, as has been
noted before, in both fluid and gyrokinetic simulations of ITG
turbulence dominated transport, without taking into account the
poloidal impurity asymmetries \cite{ap,fulopnordman}. The situation is
similar also in case of up-down asymmetric impurity populations, while
the peaking factor increases slightly in case of outboard
accumulation. 
However, in agreement with the
conclusion of \cite{poloidal}, the black dotted line in
Fig.~\ref{zscan}a shows that impurities experience outward convection
(corresponding to negative peaking factor) if the impurity density is
accumulated on the inboard side. The change in the peaking factor
becomes stronger as the asymmetry is increased, as it is illustrated
in Fig.~\ref{zscan}b. Note, that the strength of the asymmetry is also
expected to depend on $Z$, and usually it is larger for heavy
impurities, as it was shown in Sec.~\ref{sec:poloidal}. According to
Eq.~(\ref{nzpert}), in the limit of trace impurities, the poloidal
variation of the impurity density is proportional to the charge
number.

In order to examine the importance of parallel compressibility in
determining the charge dependence of the impurity peaking factor we
separate the terms independent and proportional to $\delta_{p}$
(representing the parallel compressibility) in Eq.~(\ref{pfn}) as
$S_{n}(\theta)=S^{1}_{n}(\theta)+S^{pc}_{n}(\theta)$, where
\begin{eqnarray}
S^{1}_{n}(\theta)=\mathcal{P}(\theta)\phi(\theta)\int d^3x_ze^{-x_z^2}J^2_0[z_z(\theta)]\frac{\bar{\omega}_{*z}}{\bar{\omega}-\bar{\omega}_{Dz}(\theta)}
,\label{pfn1}
\end{eqnarray}
 \begin{eqnarray}
       S^{pc}_{n}(\theta)=-\delta_p
       \int d^3x_ze^{-x_z^2}J_0[z_z(\theta)]\mathcal M_\theta
      \frac{\bar{\omega}_{*z}\mathcal{P}(\theta)\phi(\theta)J_0[z_z(\theta)]}{\bar{\omega}-\bar{\omega}_{Dz}(\theta)}
,\label{pfnpc}
\end{eqnarray}
and in Eq.~(\ref{pfT}) as $S_{T}(\theta)=S^{1}_{T}(\theta)+S^{pc}_{T}(\theta)$, where
\begin{eqnarray}
  S^{1}_{T}(\theta)=\mathcal{P}(\theta)\phi(\theta) \int d^3x_ze^{-x_z^2}J^2_{0}[z_z(\theta)]\frac{\bar{\omega} - \bar{\omega}_{*z}(x_z^2 - 3/2)a/L_{Tz}}{\bar{\omega}-\bar{\omega}_{Dz}(\theta)}
  ,\label{pfT1}
\end{eqnarray}
\begin{eqnarray}
  \hspace{-2cm}S^{pc}_{T}(\theta)= -\delta_p\int
  d^3x_ze^{-x_z^2}J_{0}[z_z(\theta)]\mathcal M_\theta\frac{[\bar{\omega} - \bar{\omega}_{*z}(x_z^2 - 3/2)a/L_{Tz}]\mathcal{P}(\theta)\phi(\theta)J_{0}[z_z(\theta)]}{\bar{\omega}-\bar{\omega}_{Dz}(\theta)}
  .\label{pfTpc}
\end{eqnarray}
Using the above notations the peaking factor can be rewritten as:
\begin{eqnarray}
  a/L_{nz}^0=\frac{\langle\mathrm{Im}[S^{1}_{T}(\theta)\phi^\ast]\rangle+\langle\mathrm{Im}[S^{pc}_{T}(\theta)\phi^\ast]\rangle
  }{\langle\mathrm{Im}[S^{1}_{n}(\theta)\phi^\ast]\rangle + \langle\mathrm{Im}[S^{pc}_{n}(\theta)\phi^\ast]\rangle}.\label{newpeak}\end{eqnarray}

\begin{figure}[htbp]
\begin{center}
 \includegraphics[width=1\textwidth]{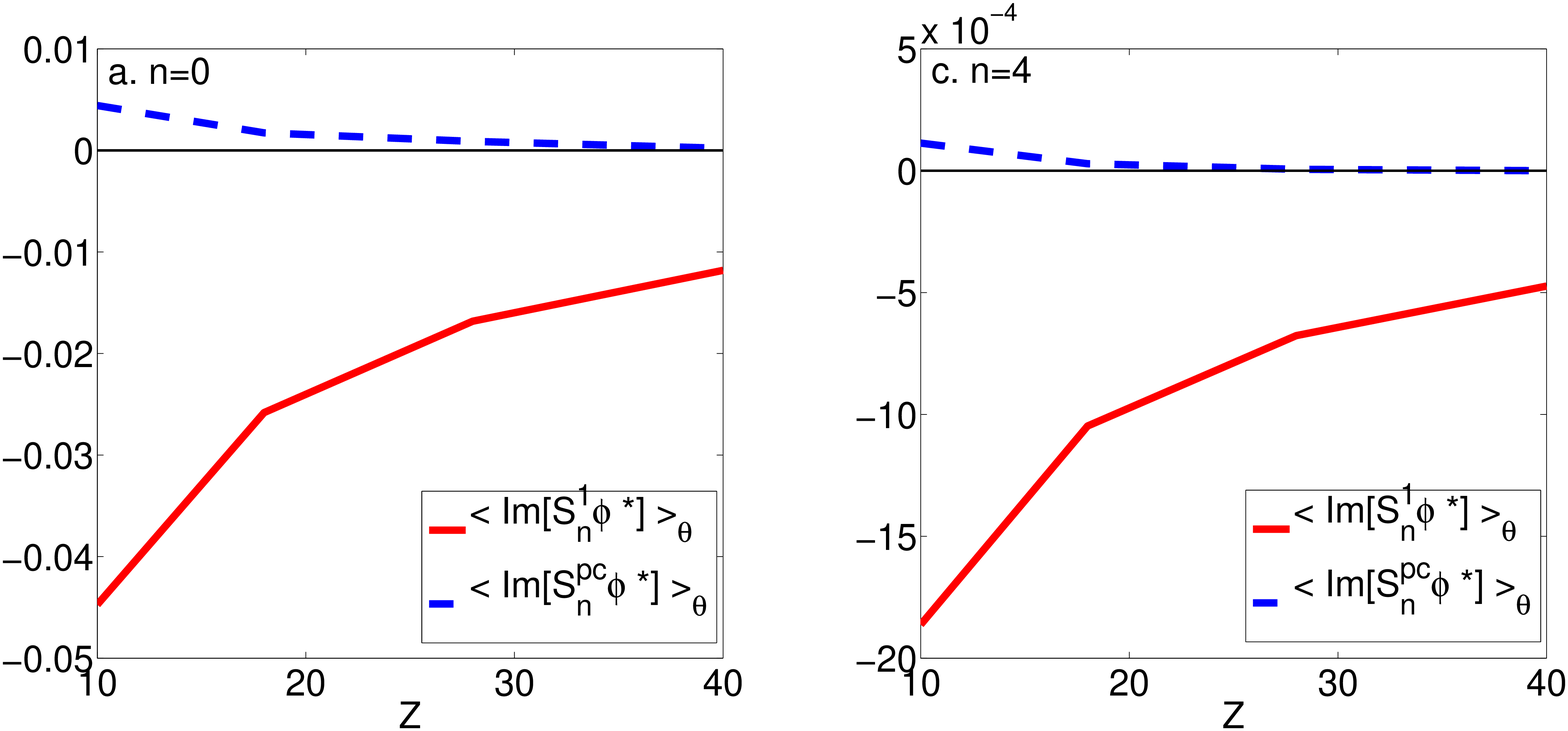} \\\includegraphics[width=1\textwidth]{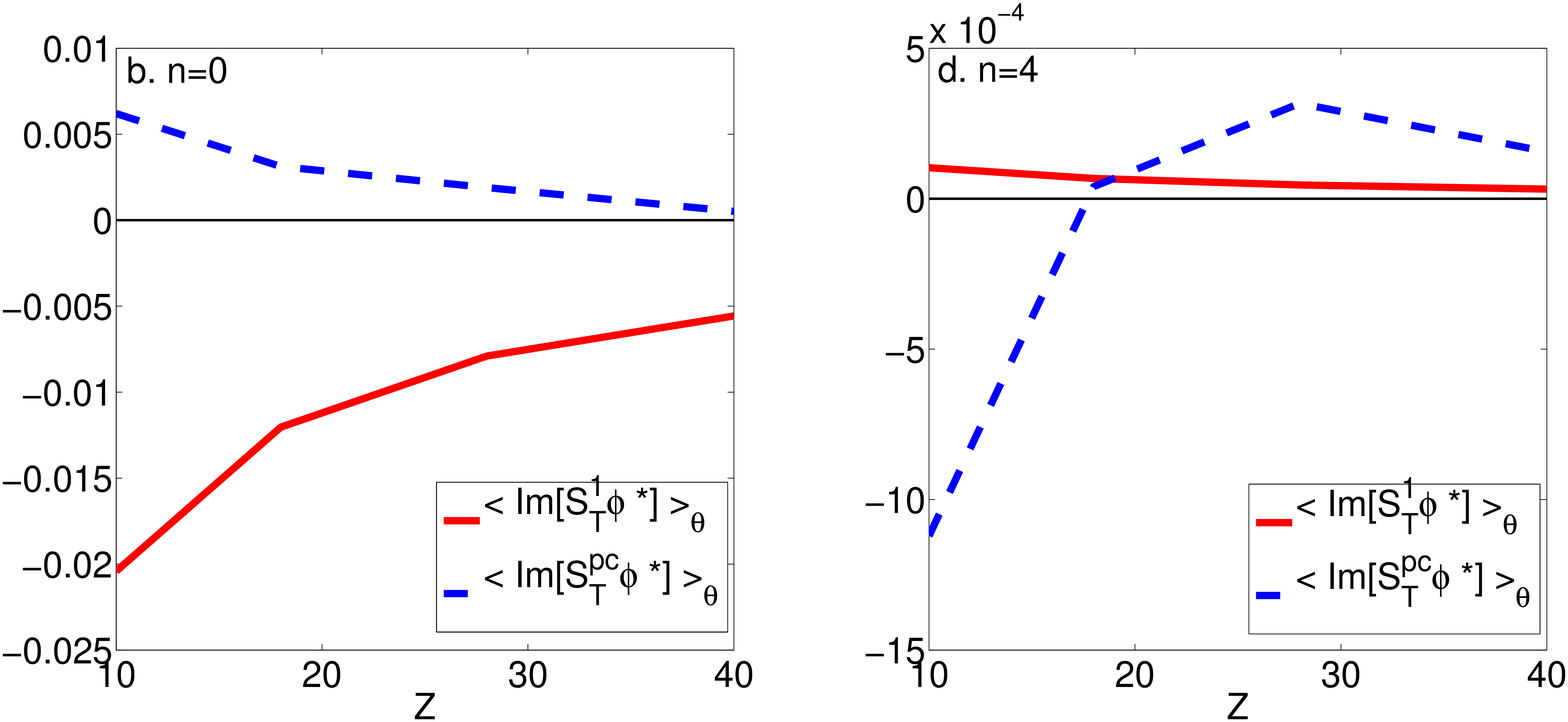}
 \caption{Terms proportional (dashed lines) and independent of
   $\delta_{p}$ (solid lines) in Eq.~(\ref{newpeak}) as functions of
   impurity charge $Z$. (a) and (c) show the terms in the denominator
   and (b) and (d) show the terms in the numerator. (a) and (b)
   correspond to the poloidally symmetric case, and (c) and (d)
   correspond to in-out asymmetry with strength $n=4$. }
\label{zscannT}
\end{center}
\end{figure}

Figure~\ref{zscannT} shows these four expressions as functions of
impurity charge for the symmetric and in-out asymmetric cases with an
asymmetry strength of $n=4$. In the denominator, the terms independent
of $\delta_{p}$, i.e. $S^{1}_{n}(\theta)$, are the dominant
contributors in both the symmetric and asymmetric cases. These are
proportional to $1/Z$. But in the numerator, the balance between the
terms proportional and independent of $\delta_{p}$ is very different
between the symmetric and asymmetric cases; in the symmetric case
$S^{1}_{T}(\theta)$ is the dominant term, while in the asymmetric case
$S^{pc}_{T}(\theta)$ is dominant. This latter term, which is due to
parallel compressibility, changes sign at moderate charge numbers
($Z\simeq 15$), which in turn leads to a sign-change in the peaking
factor.  These results show that if an in-out asymmetry is present,
the parallel compressibility effects become more important than the
other terms, and therefore, have to be taken into account. 

\begin{figure}[htbp]
\begin{center}
  \includegraphics[width=0.45\textwidth]{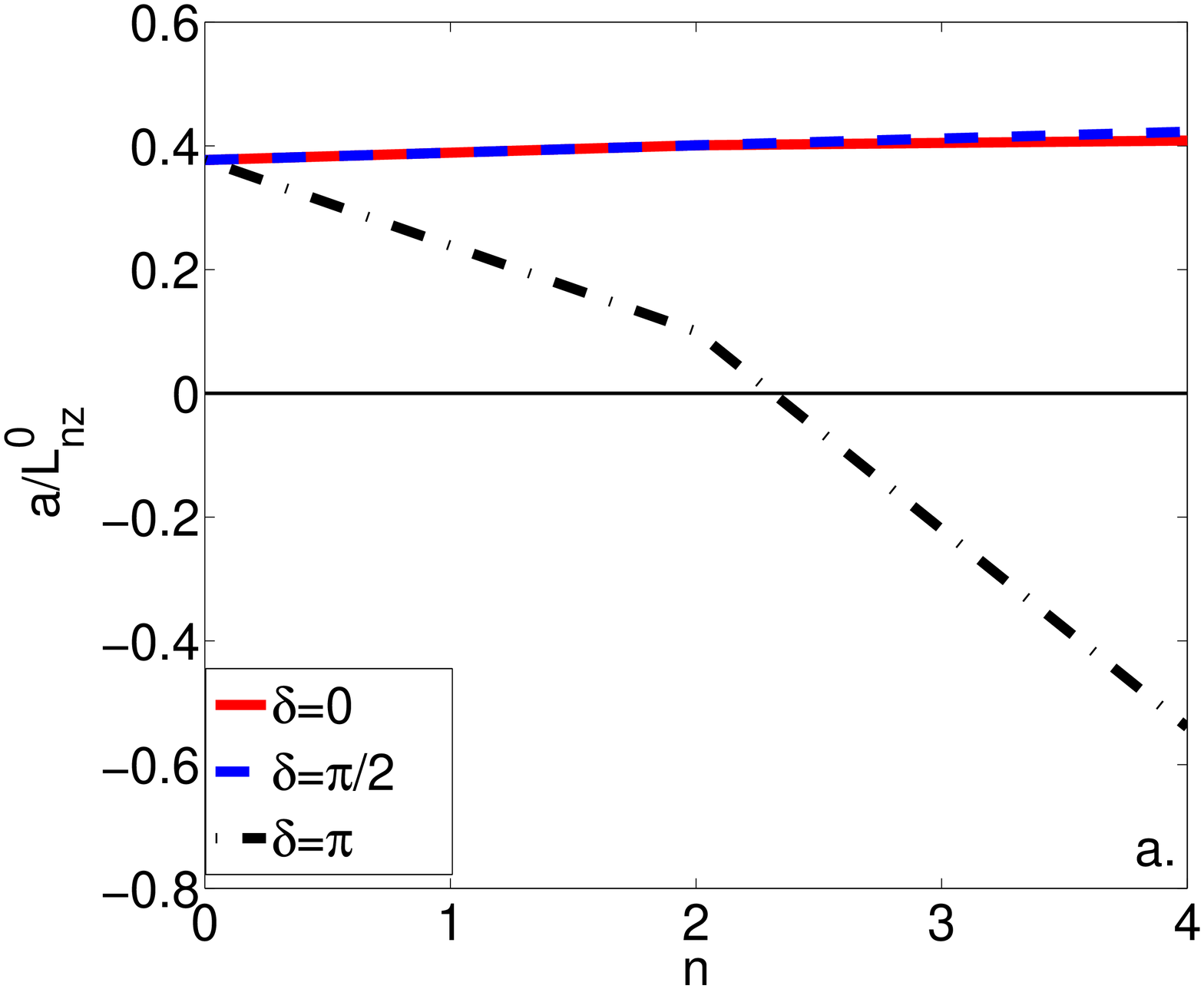}\includegraphics[width=0.45\textwidth]{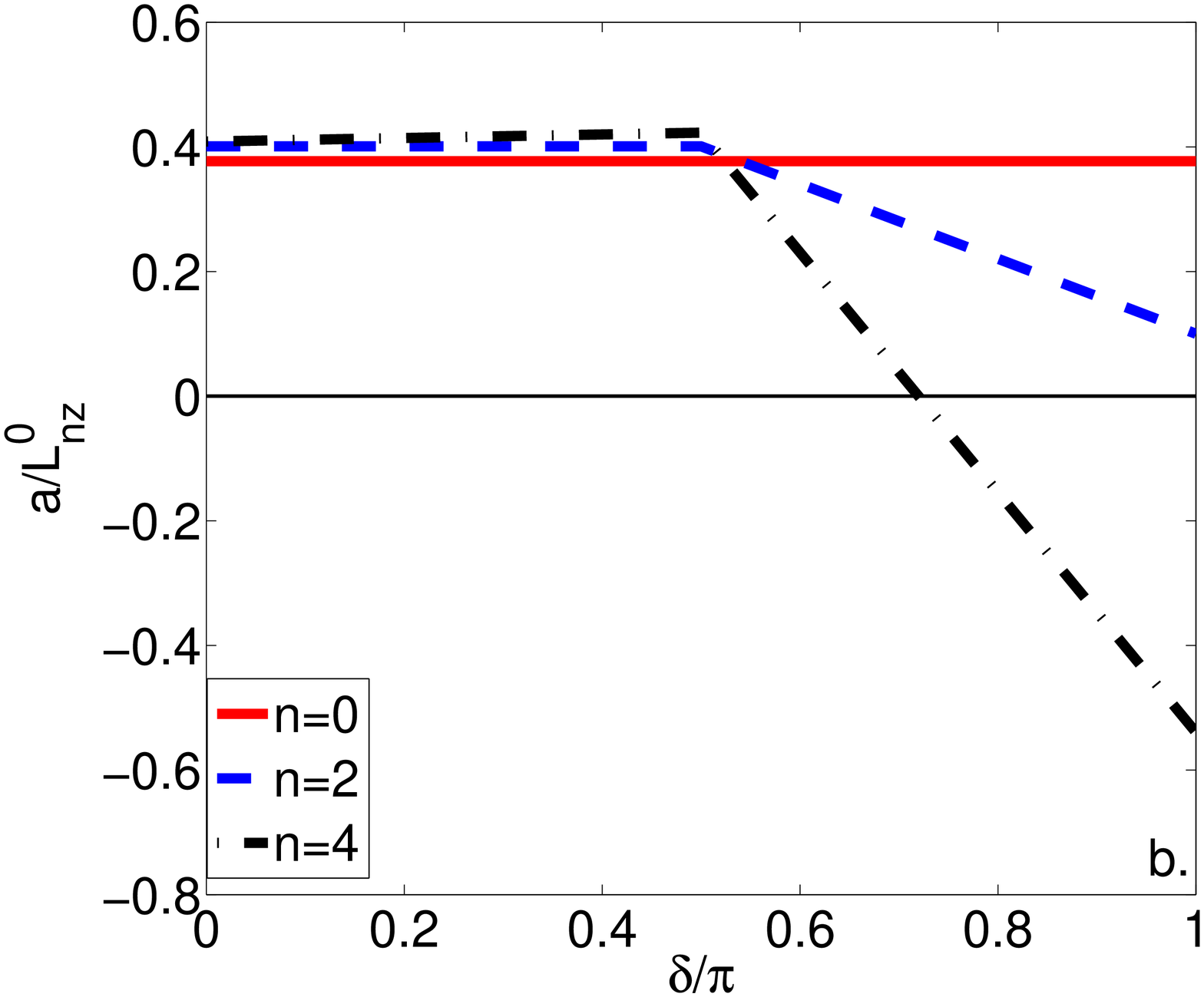}
  \caption{Peaking factor for nickel as a function of $n$ (a) and
    $\delta$ (b).  (a): symmetric impurity density (solid, red), up-down
    asymmetry (dashed, blue), and in-out asymmetry (dash-dotted, black). (b):
    symmetric impurity density (solid, red), $n=2$ (dashed, blue), $n=4$
    (dash-dotted, black).}
\label{nickel}
\end{center}
\end{figure}

In the following we will concentrate on the peaking factor for nickel
which was one of the impurities studied in Ref.~\cite{valisa}.  Figure
\ref{nickel} shows the peaking factor for nickel for various
accumulation maxima $\delta$ and asymmetry strengths $n$. The peaking
factor increases slightly for up-down accumulation but the more
dramatic change -- including a sign-change -- is expected only for
inboard accumulation. The sign change occurs when $n\simeq 2$. In
Ref.~\cite{poloidal} it was shown that without taking into account the
effect of parallel compressibility the sign change occurs when
$n\simeq 3$.  This is in agreement with the results obtained here when
parallel compressibility effects are neglected, i.e. by setting
$\delta_{p}=0$ in Eqs.~(\ref{pfn}) and (\ref{pfT}) (as will be shown
later in Fig.~\ref{nickelwoparacomp}).

Assuming that by increasing the RF heating and therefore the
temperature gradients, the asymmetry will increase as well, e.g. $n
\propto a/L_{T}$ the results shown in Fig.~\ref{nickel}a can be linked
to the experimental results illustrated in Figs.~6 and 7 in
Ref.~\cite{valisa}. In this reference the experimental observation of
the effect of ICRH heating on the nickel transport in the plasma core
at JET is discussed. It is shown that the application of the ICRH
leads to an outward impurity flux (negative peaking factor). As the
applied heating is increased and therefore, the temperature gradient
peaked, the outward flux is further increased leading to more negative
peaking factors.

\subsection{Temperature gradient dependence}
\begin{figure}[htbp]
\begin{center}
  \includegraphics[width=0.45\textwidth]{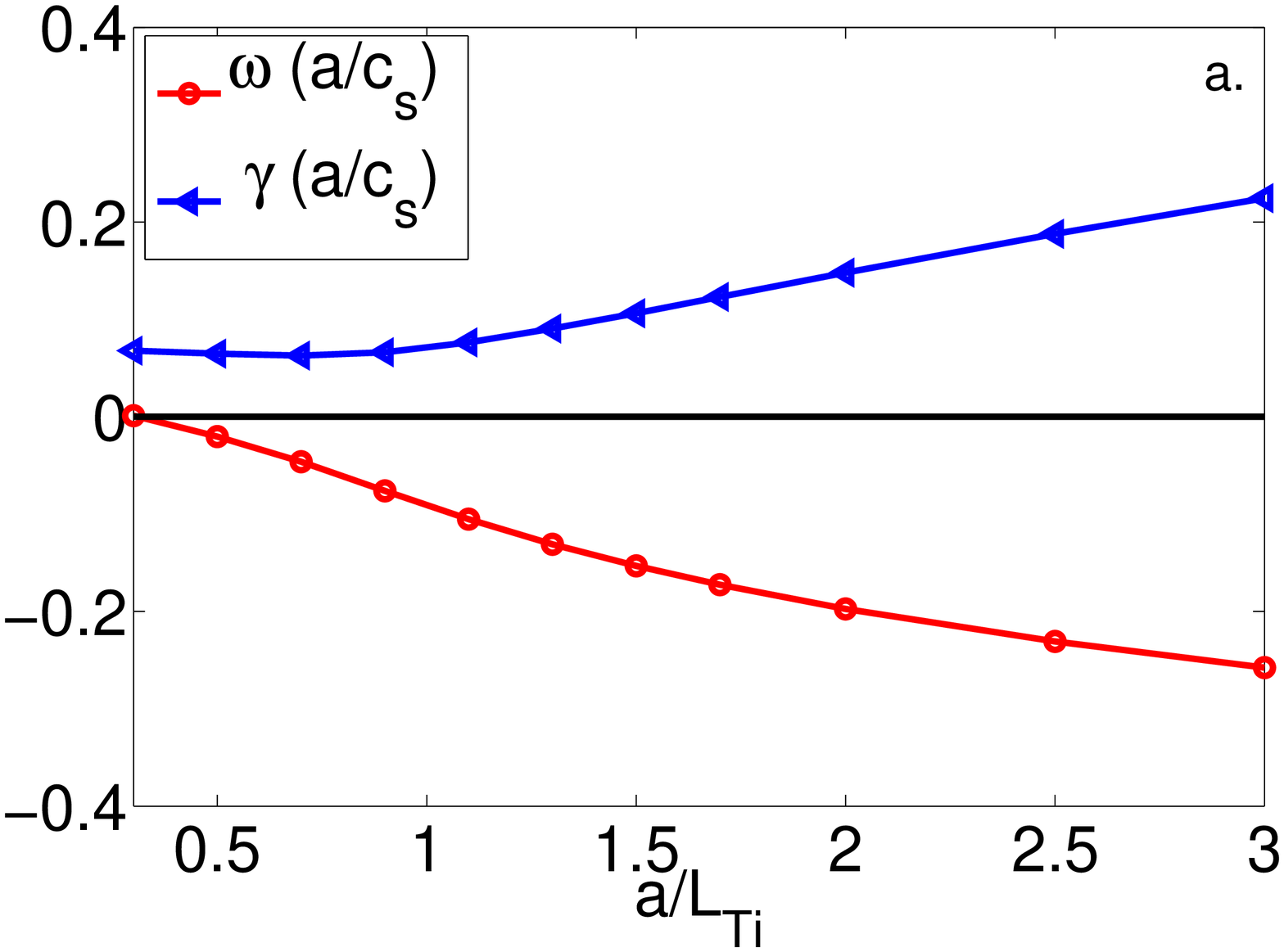}
  \includegraphics[width=0.45\textwidth]{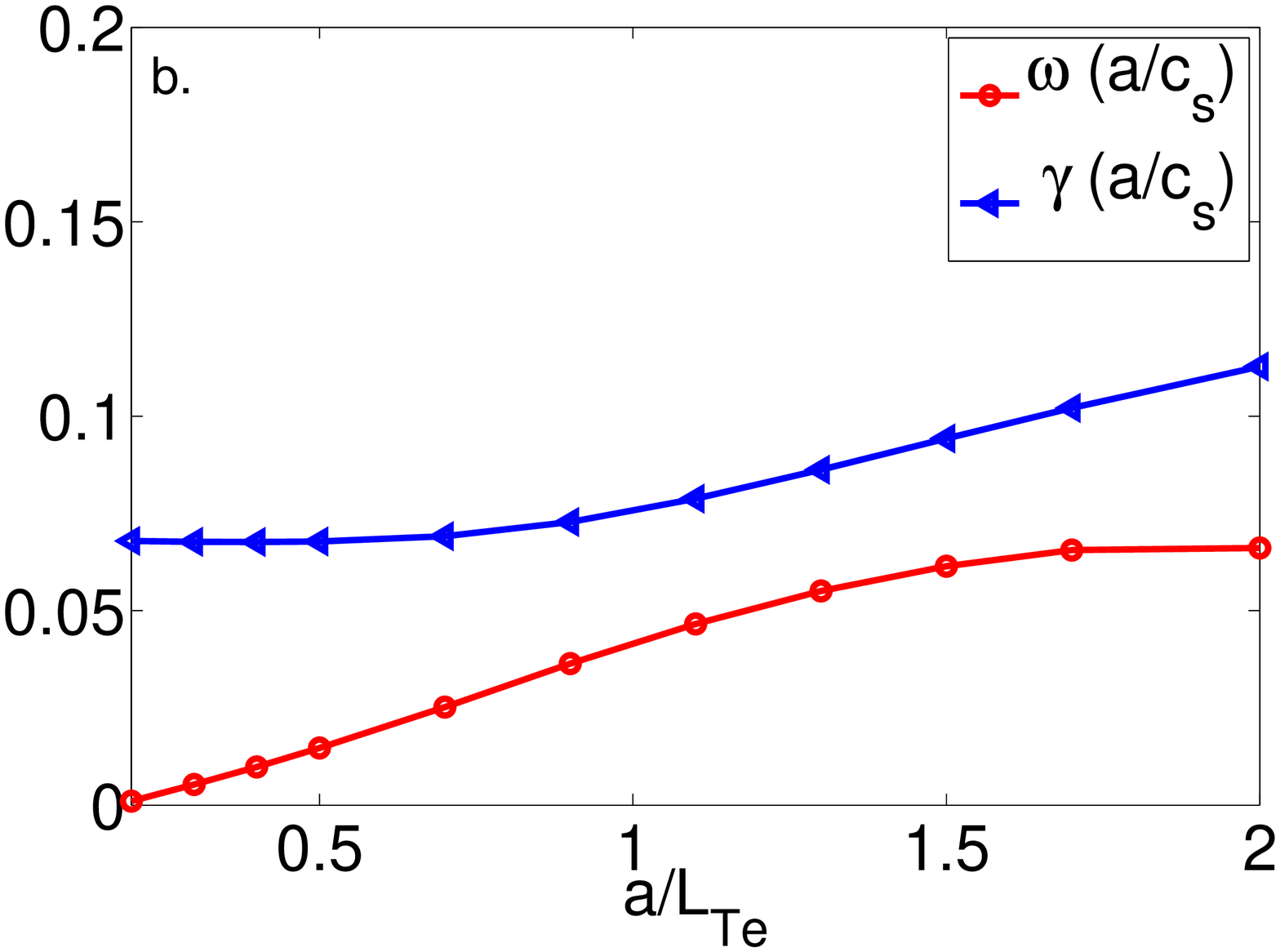} 
  \caption{Real and imaginary parts of $\omega=\omega_{r}+i\gamma$ as
    function of $a/L_{Ti}$ (a) and $a/L_{Te}$ (b) obtained by \gyro for
    the baseline case. Blue lines (with circle markers) represent the
    real part, blue lines (triangle markers) correspond to the
    imaginary part of the eigenvalue. The frequencies are normalized to
    $c_s/a$. }
\label{eigenvalues}
\end{center}
\end{figure}

\begin{figure}[htbp]
\begin{center}
  \includegraphics[width=0.45\textwidth]{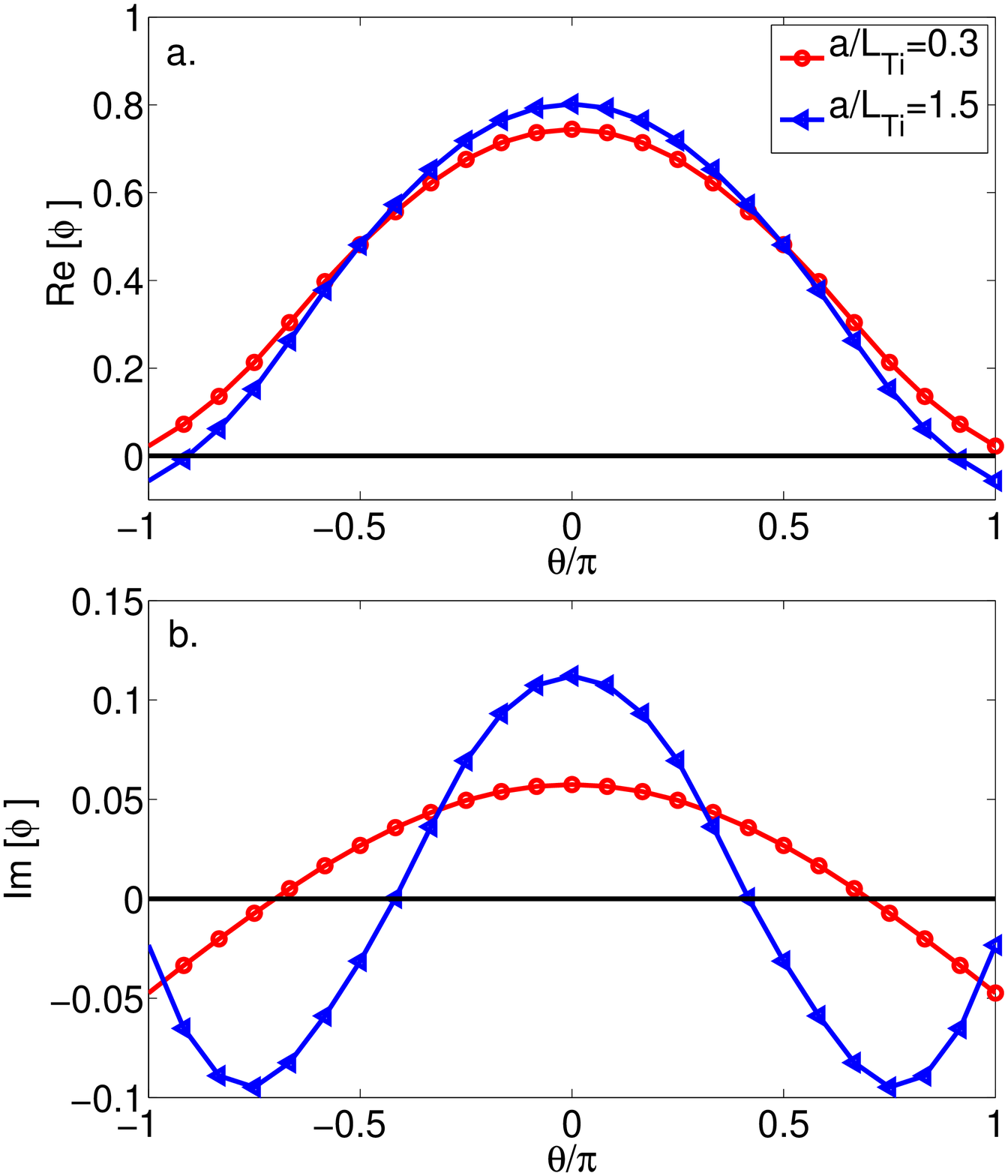}
  \includegraphics[width=0.45\textwidth]{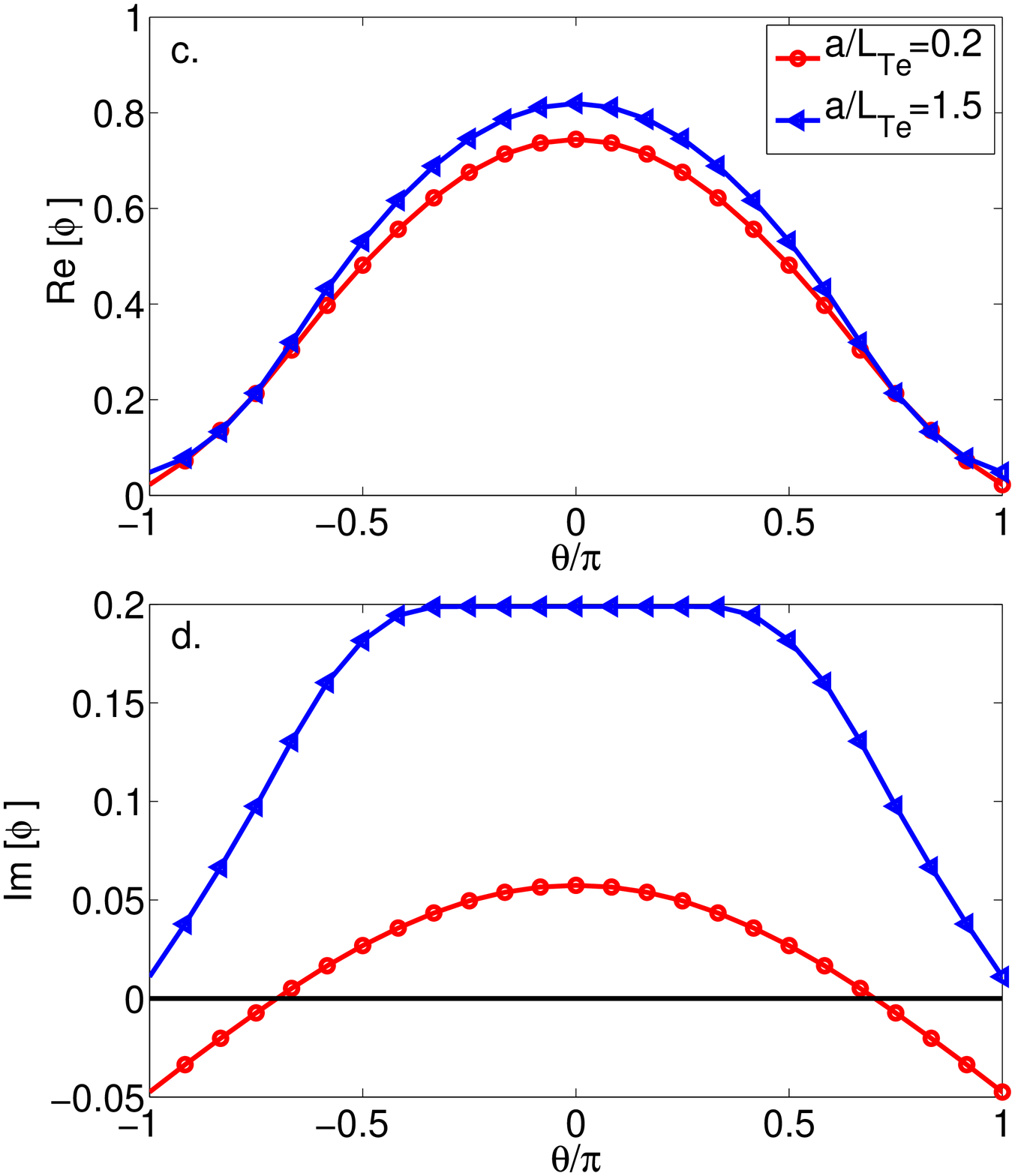} 
  \caption{Real and imaginary parts of the electrostatic potentials in
    the cases corresponding to Fig.~\ref{eigenvalues} for different
    ion- (a,b) and electron (c,d) temperature gradients. (a) and (c)
    show the real parts and (b) and (d) show the imaginary parts of
    $\phi$ (note the different scale).  Red lines (circle markers)
    show the temperature gradient corresponding to the baseline case
    and blue lines (triangle markers) show the case with the larger
    temperature gradient.   }
\label{potentials}
\end{center}
\end{figure}

The eigenvalues and electrostatic potentials as functions of ion- and
electron temperature gradients are shown in
Figs.~\ref{eigenvalues}-\ref{potentials}. 
As expected, if we increase the ion temperature gradient, the
turbulence becomes more ITG-dominated (the real part of the mode
frequency $\omega_r$ is negative), while if we increase the electron
temperature gradient, TE-mode driven turbulence will dominate
($\omega_r$ is positive).  The shape of the imaginary part of the
potential $\mathrm{Im}[\phi]$ varies strongly by increasing the
temperature gradient for both the ITG and TE-mode dominated cases,
while the real parts of the potential $\mathrm{Re}[\phi]$ are not
modified significantly, see
Fig.~\ref{potentials}. 
The imaginary part of the potential plays an important role in the
parallel compressibility terms and therefore the change in the
temperature gradient will modify the impurity peaking factor
considerably.

Previous works highlighted the difference in peaking factors in ITG
and TE dominated cases, and concluded that ITG-dominated turbulence
will always generate inward pinch of impurities, while in TE-mode
driven turbulence outward convection in the plasma core (for
$r/a\simeq 0.2$) can be expected. Both linear \cite{angionipop07,ap}
and non-linear \cite{angioniNF09} gyrokinetic simulations have shown
that the latter is due to the contribution from the parallel dynamics
which can reverse the impurity convection from inwards to outwards for
modes propagating in the electron direction. These results are in
agreement, under some conditions, with the experimental observation
that the impurity convection changes sign from inward to outward when
a strong central peaking of the electron temperature arises as a
response to strong localized central electron heating. However, these
results cannot explain the outward convection of impurities observed
in experiments with RF heating where the ITG-mode is the dominant
instability.

Our results show, that the peaking factor is negative in both ITG- and
TE-dominated cases if the impurities accumulate on the inboard
side. However, the peaking factor is influenced very strongly by
increasing temperature gradients and will become positive again above
a certain temperature gradient.
 
\begin{figure}[htbp]
  \begin{center}
  \includegraphics[width=0.45\textwidth]{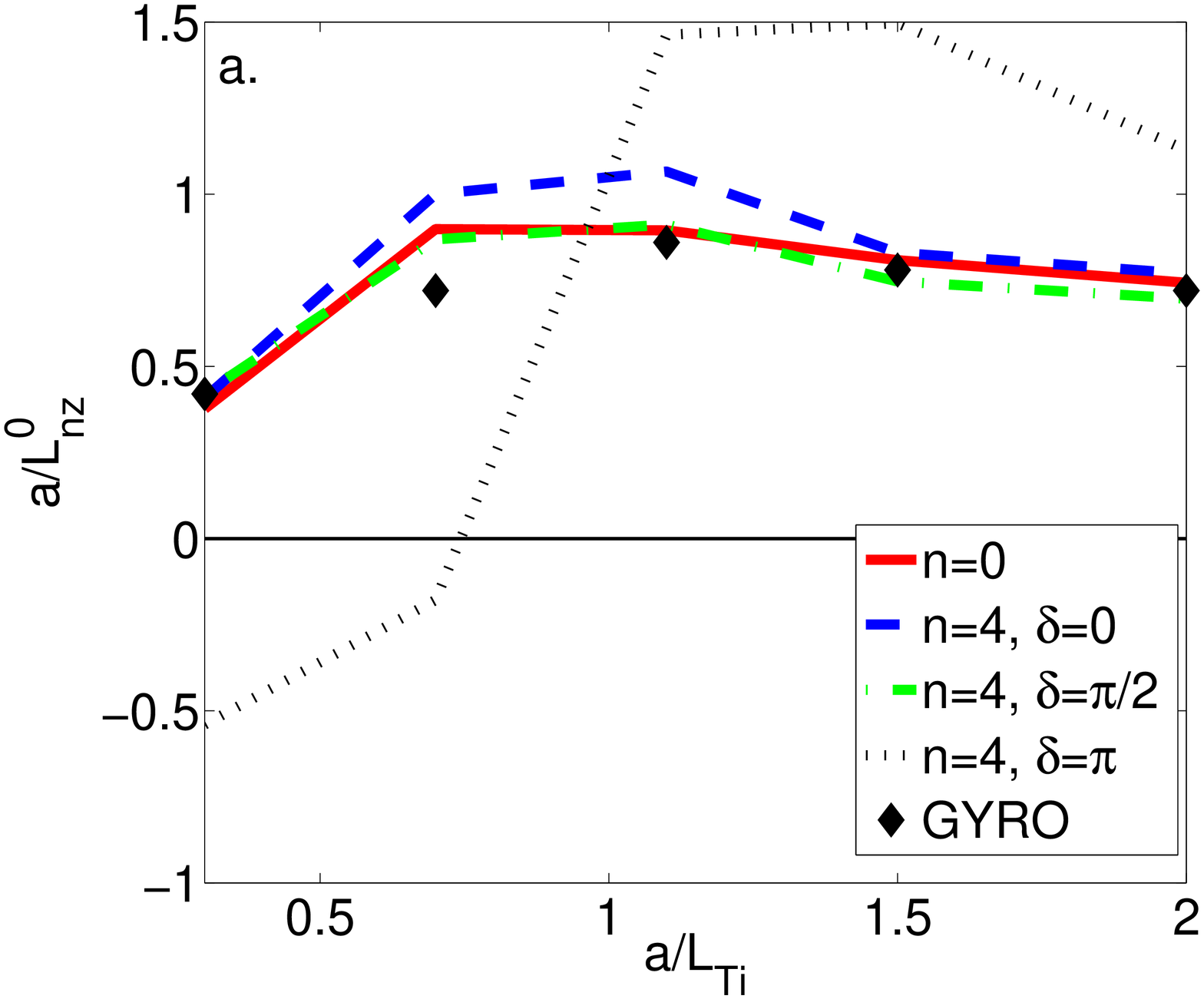}  \includegraphics[width=0.45\textwidth]{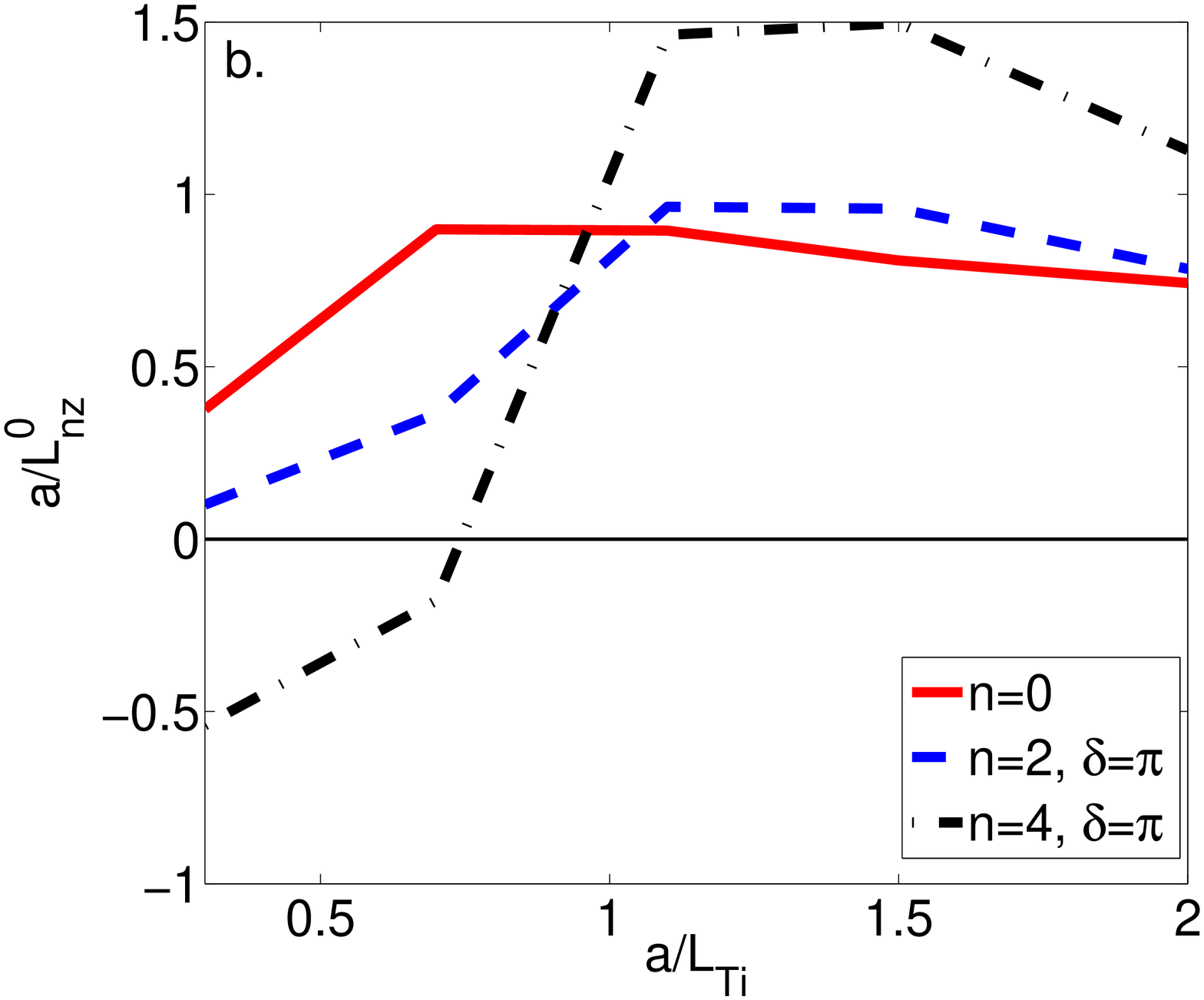}
\caption{ Peaking factor for nickel as a function of ion temperature
  gradient for different values of $\delta$ (a) and $n$ (b). In both
  figures the solid line represents the case of poloidally symmetric
  impurity distribution; in figure (a) this case is compared to \gyro
  simulations (black diamonds). (a): $n=4$ -- out-in asymmetry
  (dashed), up-down asymmetry (dash-dotted), in-out asymmetry
  (dotted). (b): in-out asymmetry -- $n=2$ (dashed), $n=4$
  (dash-dotted).}
\label{tgradscan1}
\end{center}
\end{figure} 

Figure \ref{tgradscan1} shows the ion temperature-gradient scaling of
the peaking factor and the diamond symbols show these values obtained
by \gyro simulations in the poloidally symmetric limit, which present
good agreement with our results.  The peaking factor is not very
sensitive to the ion temperature gradient as long as the impurity
density is poloidally symmetric or if it is up-down or out-in
asymmetric. However, if the impurity density is in-out asymmetric then the sign
of the peaking factor is negative for low temperature gradients and
positive (and even larger than in the poloidally symmetric case) for
large temperature gradients. The threshold for the sign change for the
experimental scenario studied in this paper is $a/L_{Ti}\simeq 1$.
As the strength of the asymmetry, $n$, is increased for an
in-out asymmetry the modifications of the peaking factor becomes
stronger, see Fig.~\ref{tgradscan1}b.  
 
\begin{figure}[htbp]
  \begin{center}
  \includegraphics[width=0.45\textwidth]{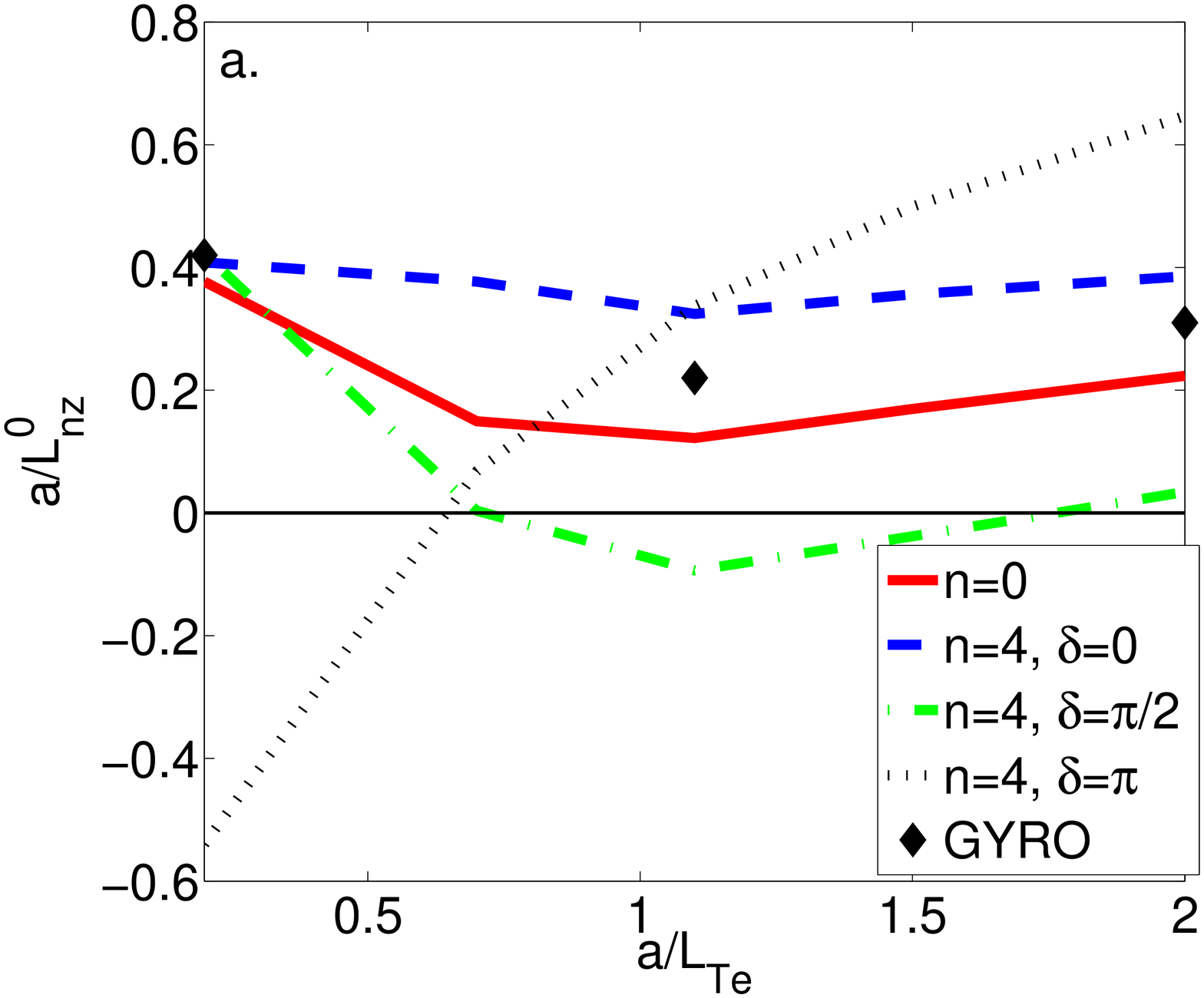}  \includegraphics[width=0.45\textwidth]{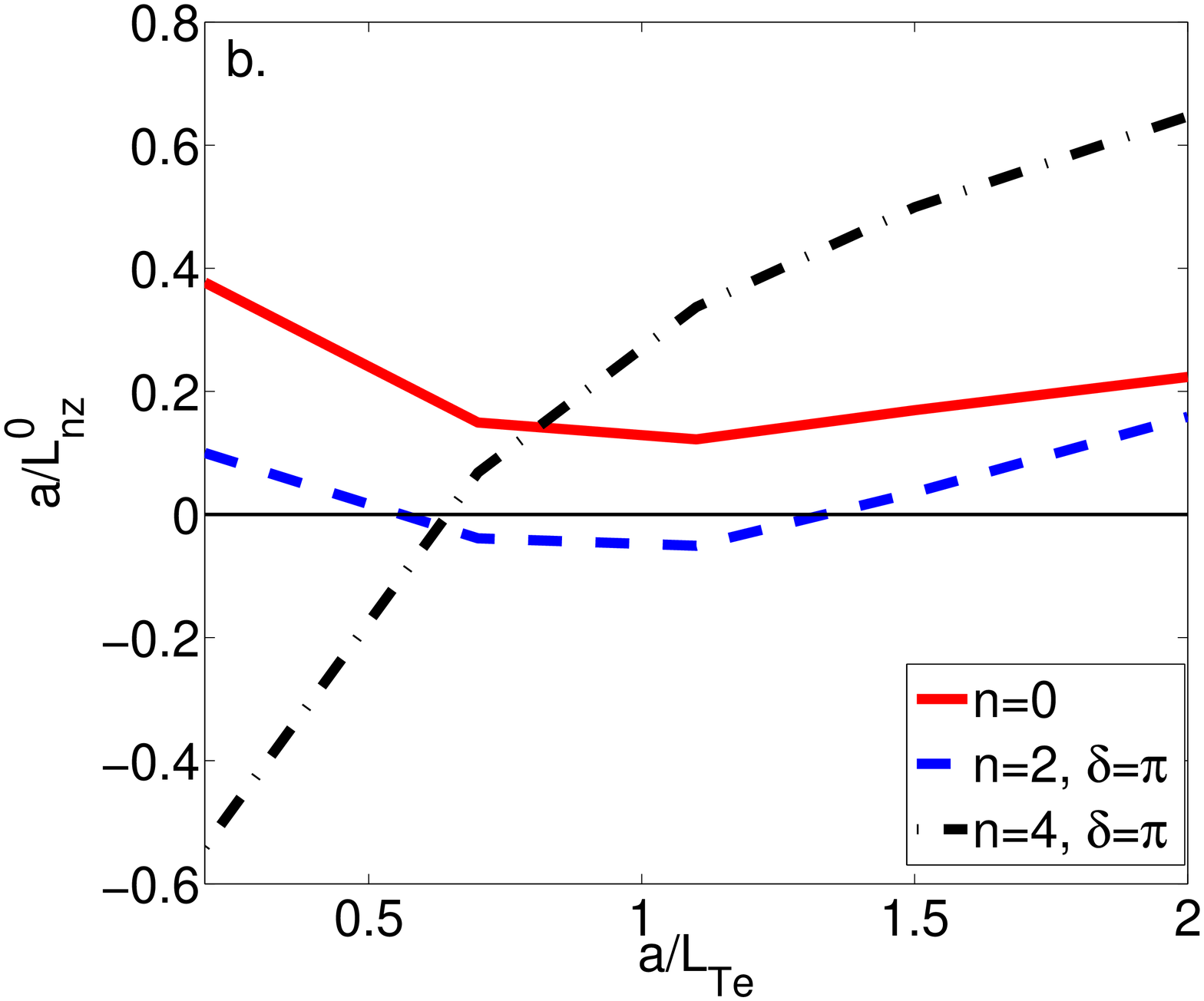}
  \caption{ Peaking factor for nickel as a function of electron
    temperature gradient for different values of $\delta$ (a) and $n$
    (b). In both figures the solid line represents the case of
    poloidally symmetric impurity distribution; in figure (a) this
    case is compared to \gyro simulations (black diamonds). (a): $n=4$
    -- out-in asymmetry (dashed, blue), up-down asymmetry
    (dash-dotted, green),
    in-out asymmetry (dotted, black). (b): in-out asymmetry -- $n=2$
    (dashed, blue), $n=4$ (dash-dotted, black).  }
\label{tgradscan2}
\end{center}
\end{figure}

As the electron temperature gradient is increased the turbulence
becomes more TE dominated. Figure \ref{tgradscan2} shows that the
absolute value of the peaking factor is reduced up to a certain
temperature gradient and then starts to increase again, regardless of
the sign of the asymmetry. The inboard accumulation gives negative
peaking factors also in this case. One of the most interesting
differences compared to the ITG-dominated case is that also an up-down
asymmetry can lead to a sign change for $a/L_{Te}\simeq 0.5$. This can
be a contributing reason for the observed flat impurity density
profiles in plasmas with electron cyclotron resonance heating (ECRH),
where the in-out accumulation mechanism described in
\cite{ingessonppcf} and in Sec.~\ref{sec:poloidal} would not be
present. Up-down asymmetries have been observed in electron cyclotron
(EC) heated plasmas \cite{condrea}. The physical mechanism for these
up-down asymmetries was attributed to neoclassical effects, but since
the magnitude of the asymmetries reported in Ref.~\cite{condrea} was
very large it is possible that there are other reasons that may be
related to the ECRH as well.

\subsection{Density gradient   dependence }
\begin{figure}[htbp] 
\begin{center}
 \includegraphics[width=0.45\textwidth]{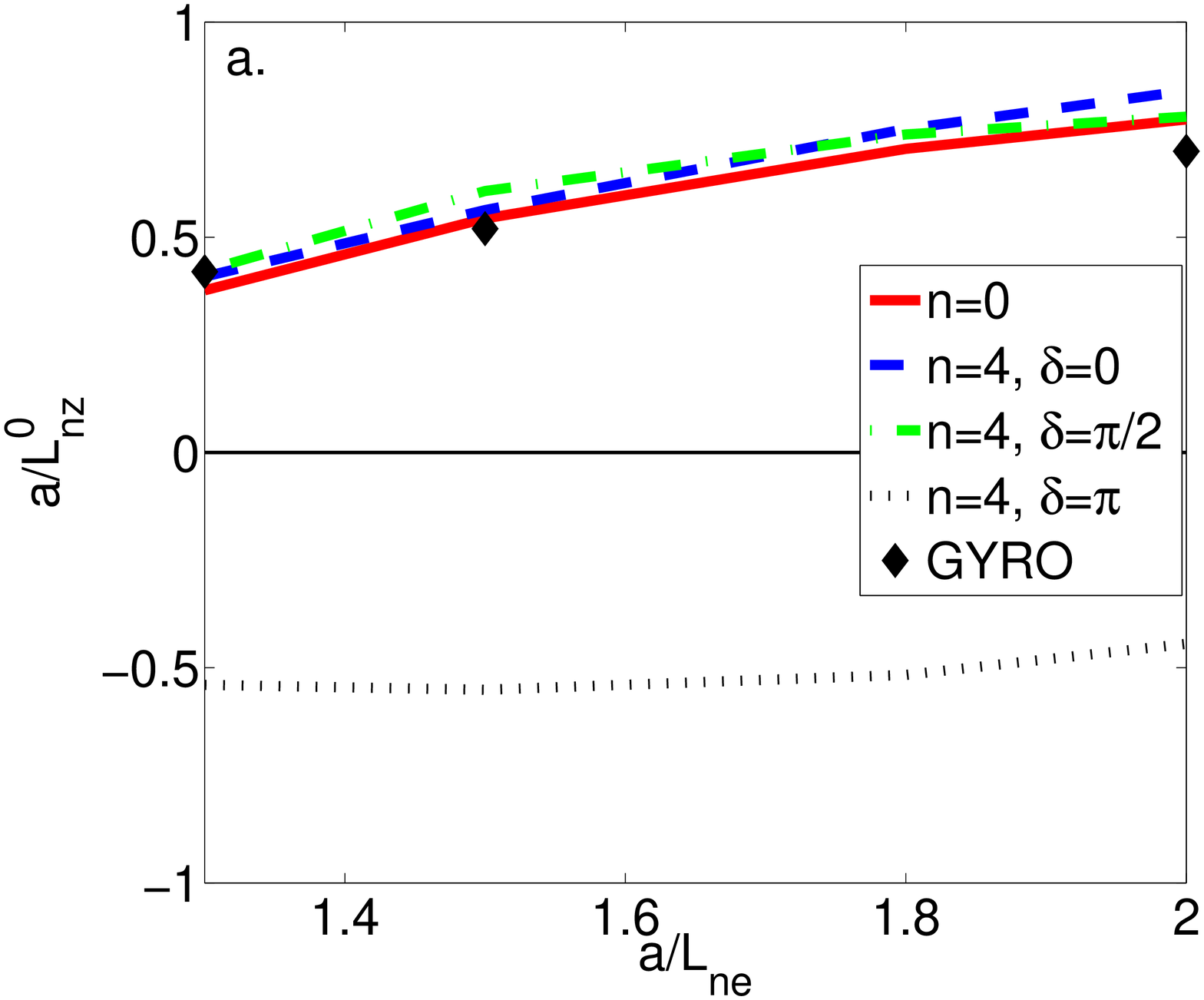} \includegraphics[width=0.45\textwidth]{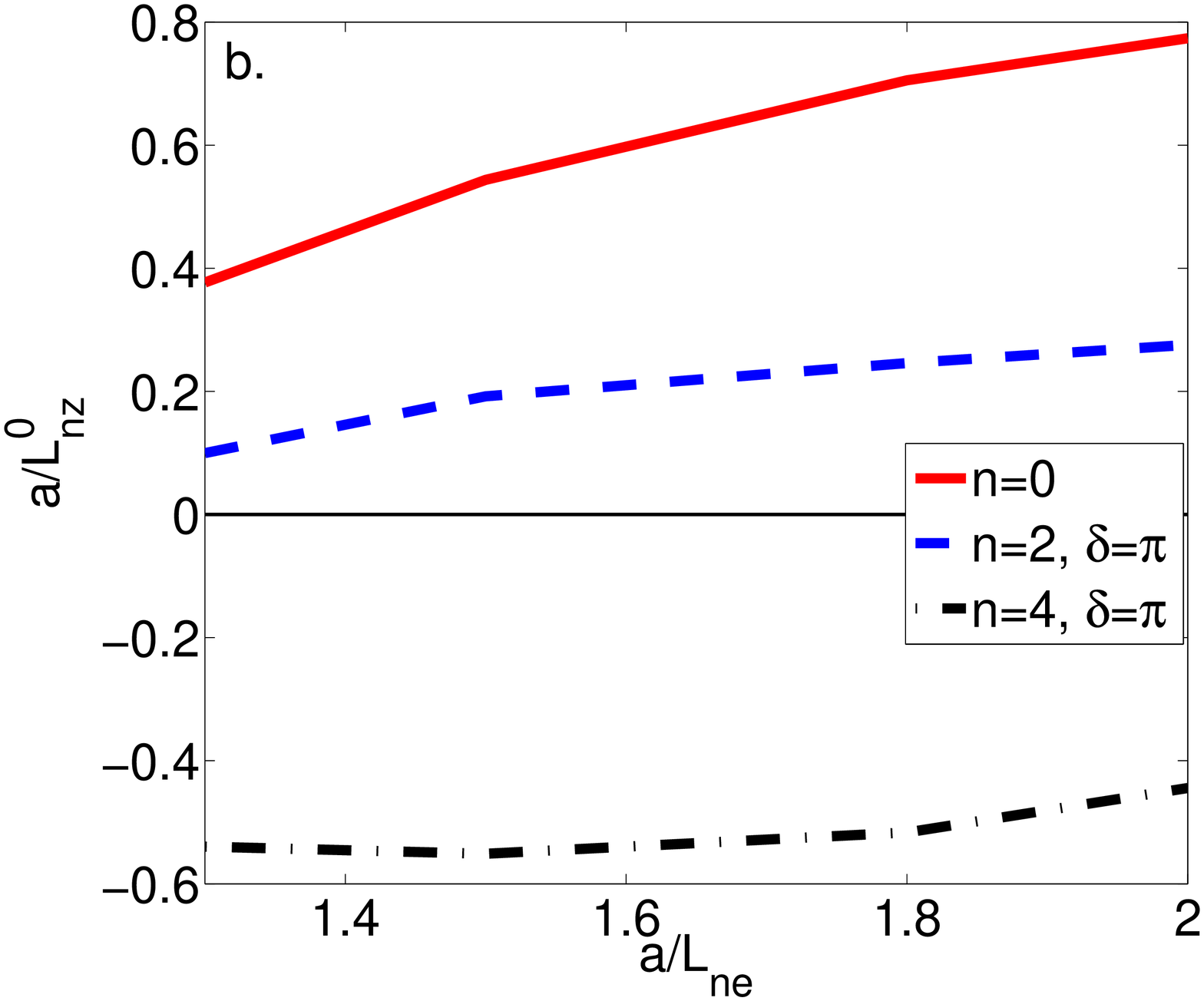}
 \caption{ Peaking factor for nickel as a function of electron density
   gradient for different values of $\delta$ (a) and $n$ (b). In both
   figures the solid line represents the case of poloidally symmetric
   impurity distribution; in figure (a) this case is compared to \gyro
   simulations (black diamonds). (a): $n=4$ -- out-in asymmetry
   (dashed, blue), up-down asymmetry (dash-dotted, green), in-out asymmetry
   (dotted, black). (b): in-out asymmetry -- $n=2$ (dashed, blue), $n=4$
   (dash-dotted, black).}
\label{ngradscan}
\end{center}
\end{figure}

\begin{figure}[htbp]
\begin{center}
  \includegraphics[width=0.45\textwidth]{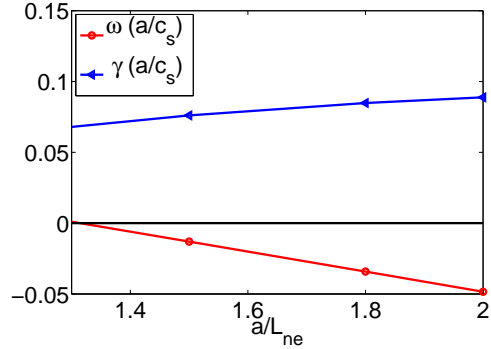}
  \caption{Real and imaginary parts of the eigenvalues as function of
    $a/L_{ne}$ obtained by \gyro$\!$. Red lines
    (with circle markers) represent the real part, blue lines
    (triangle markers) correspond to the imaginary part of the
    eigenvalue. The frequencies are normalized to $c_s/a$.}
\label{eiglne}
\end{center}
\end{figure}

\begin{figure}[htbp]
\begin{center}
  \includegraphics[width=0.45\textwidth]{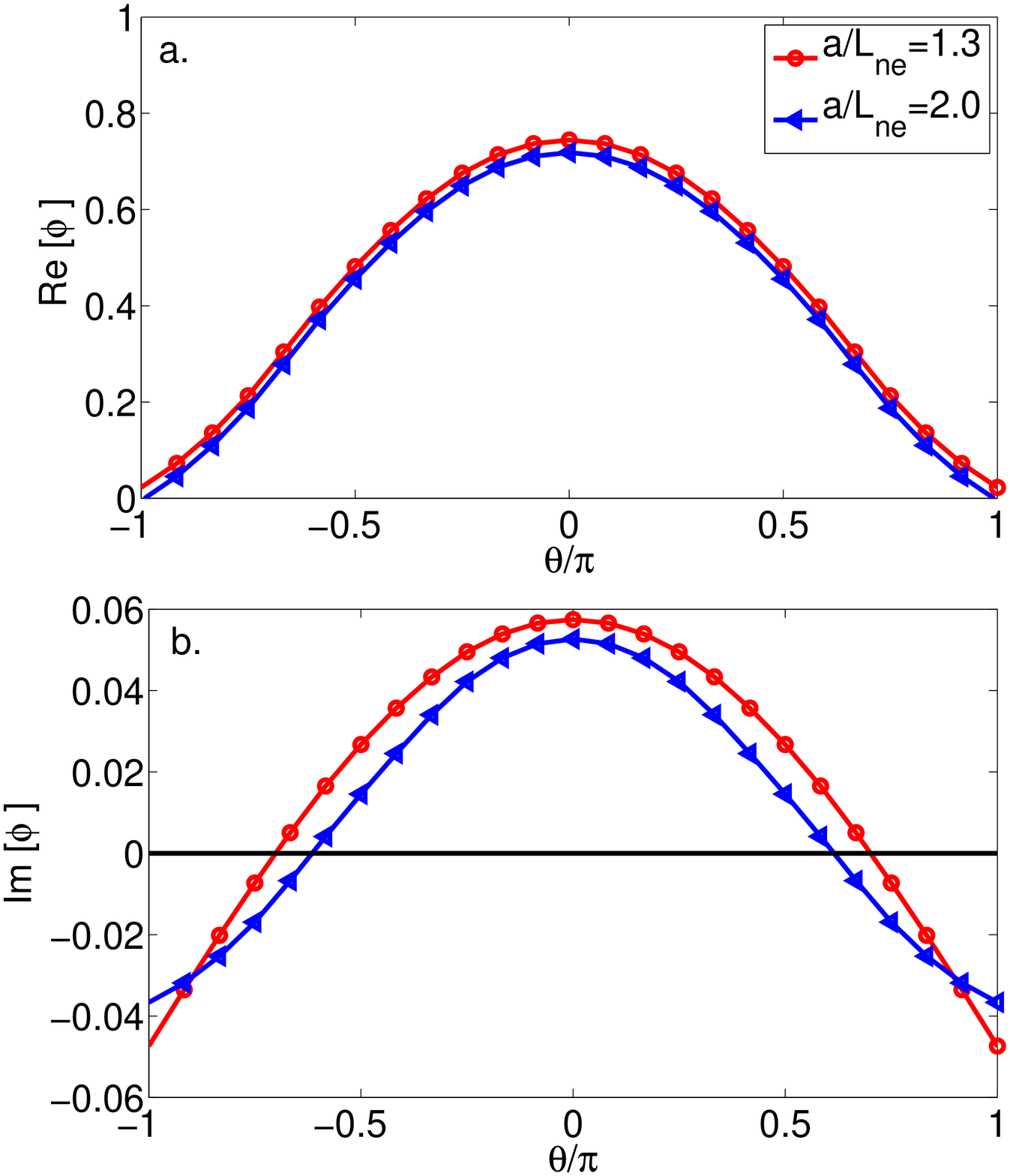}
  \caption{Real- (a) and imaginary (b) parts of the electrostatic potentials
    for two different density gradients.}
\label{potentialslne}
\end{center}
\end{figure}

The density gradient scaling of the peaking factor is shown in
Fig.~\ref{ngradscan}. The peaking factor is slightly increasing in the
case of poloidal symmetry and in the cases of up-down or out-in
asymmetries, as the electron density peaking is increased.  Also here,
the in-out asymmetric impurity density leads to negative peaking
factor if the strength of the asymmetry is sufficient.  The peaking
factor is fairly insensitive to the density gradient in this case. It
is interesting to note, that  even though the ITG-mode is the dominant
instability here, (see Fig.~\ref{eiglne}), the peaking factor remains
negative as the ITG-mode is becoming stronger in the case of an in-out
asymmetry. This can be contrasted to the case when the stronger
ion-temperature gradient led to a sign change and positive peaking
factor, see Fig.~\ref{tgradscan1}a. The difference between the two
cases can be understood by comparing the shape of the imaginary part
of the electrostatic potential $\mathrm{Im}[\phi]$ in
Fig.~\ref{potentialslne}b with that in Fig.~\ref{potentials}b. It can
be noted that by changing the temperature gradient,
$\mathrm{Im}[\phi]$ changes significantly, while changing the density
profile does not result in a significant difference. Note that it is
mainly the part of the potential which is close to $\theta=\pi$ which
is important, and that is considerably different for the two
temperature gradients shown in Fig.~\ref{potentials}b and therefore
also the result for the peaking factor changes dramatically. The
difference in the electrostatic potentials is the underlying reason
for the difference in the trends seen in Figs.~\ref{tgradscan1} and
\ref{ngradscan} as the ITG-mode become stronger by increasing the
$a/L_{Ti}$ or $a/L_{ne}$, respectively.


\subsection{Effect of parallel compressibility}

\begin{figure}[htbp]
  \begin{center}
    \includegraphics[width=0.45\textwidth]{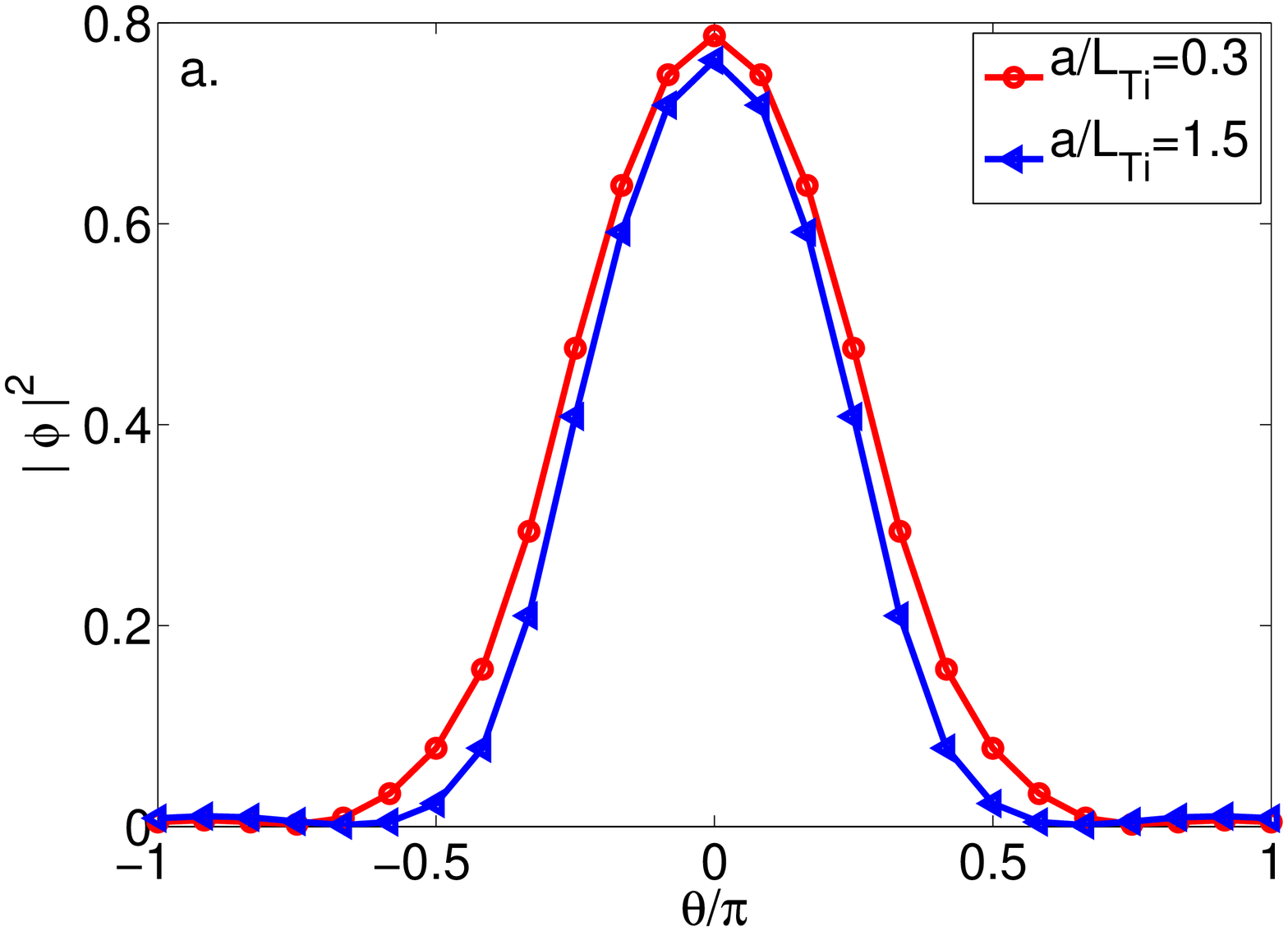}\includegraphics[width=0.45\textwidth]{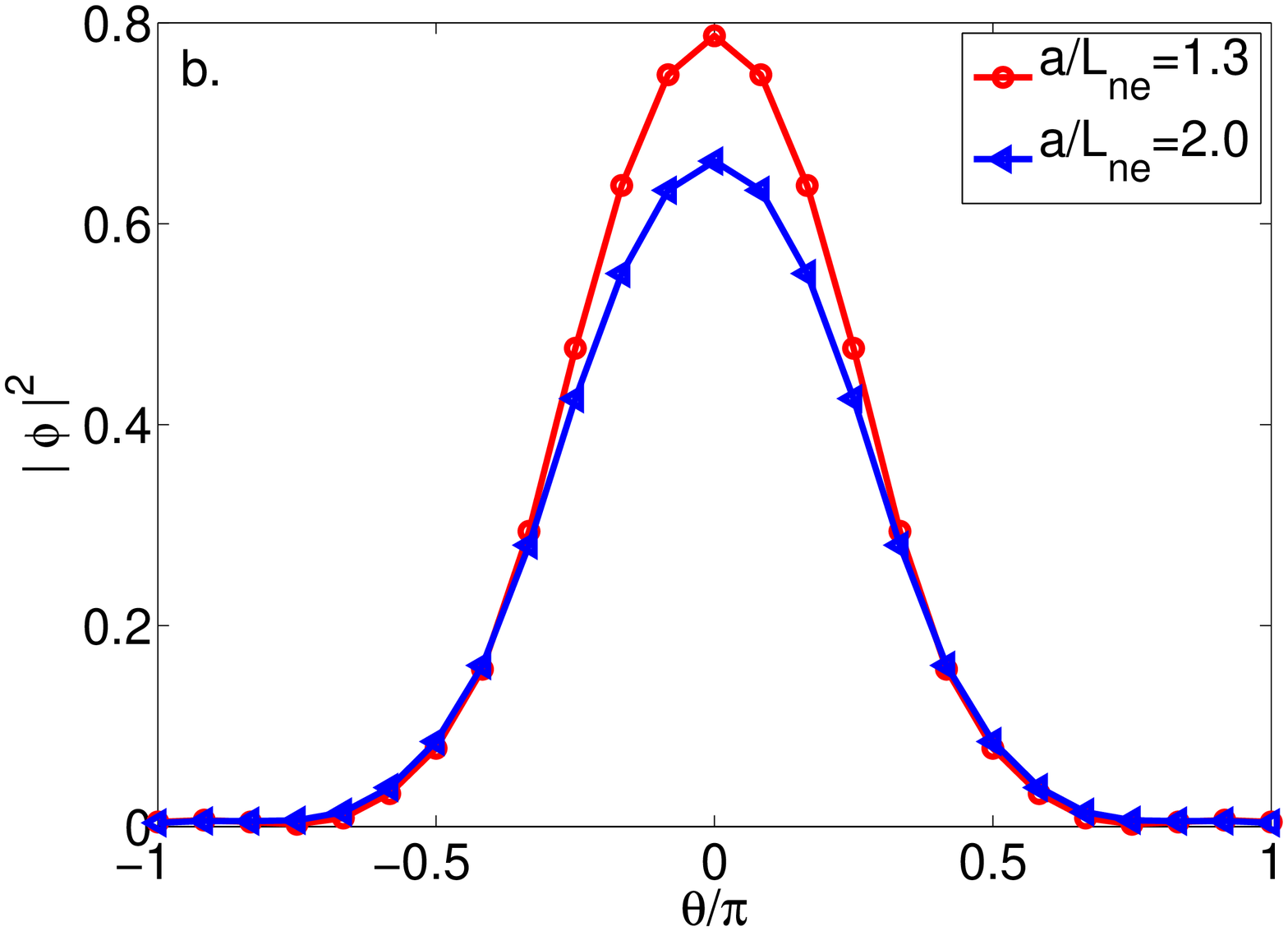}\caption{The
      square of the absolute value of the electrostatic potentials,
      $|\phi|^2$ for different ion temperature gradients and density
      gradients.}
\label{phi2woparcomp}
\end{center}
\end{figure}

\begin{figure}[htbp]
\begin{center}
\includegraphics[width=0.45\textwidth]{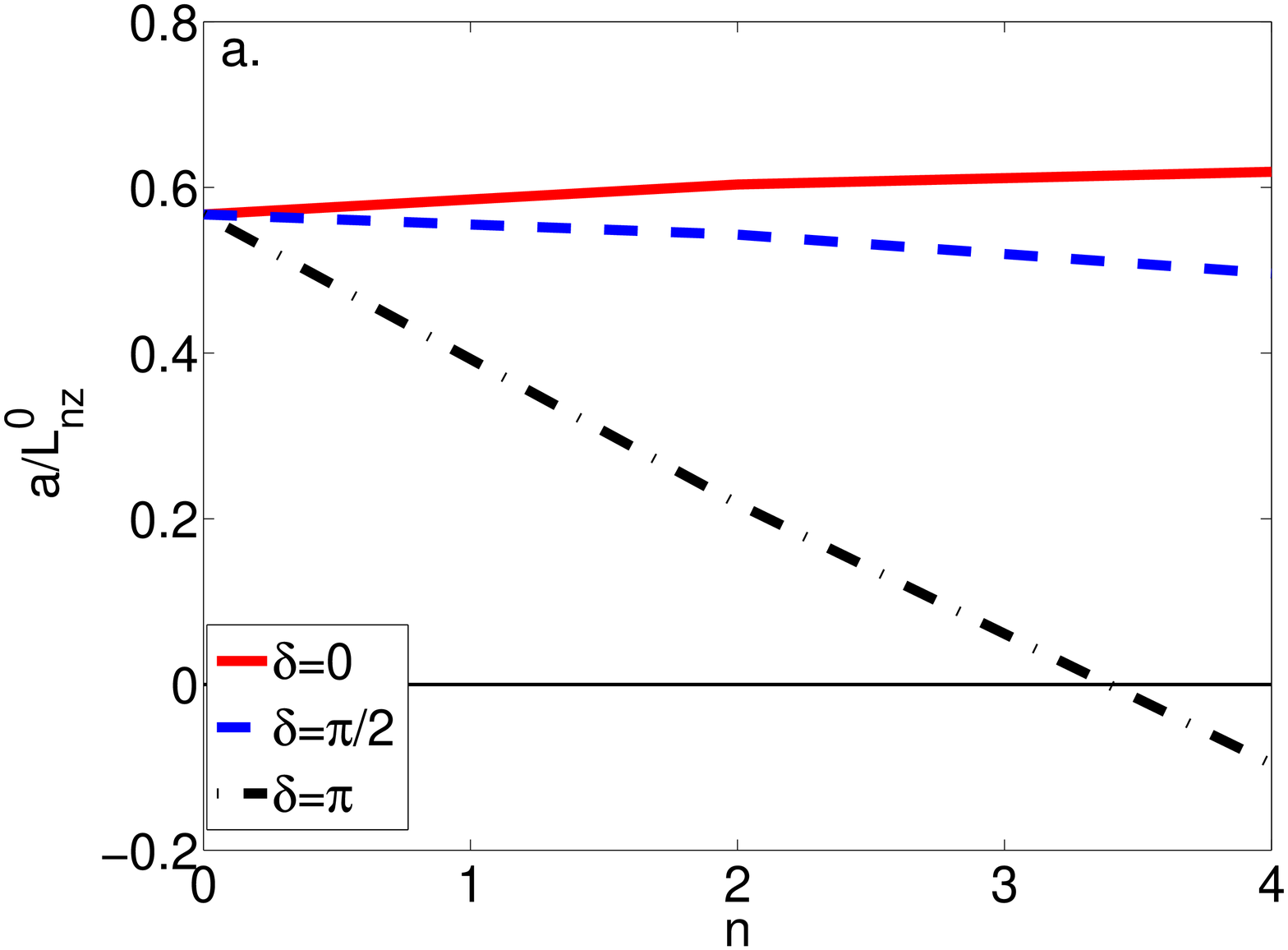}\includegraphics[width=0.45\textwidth]{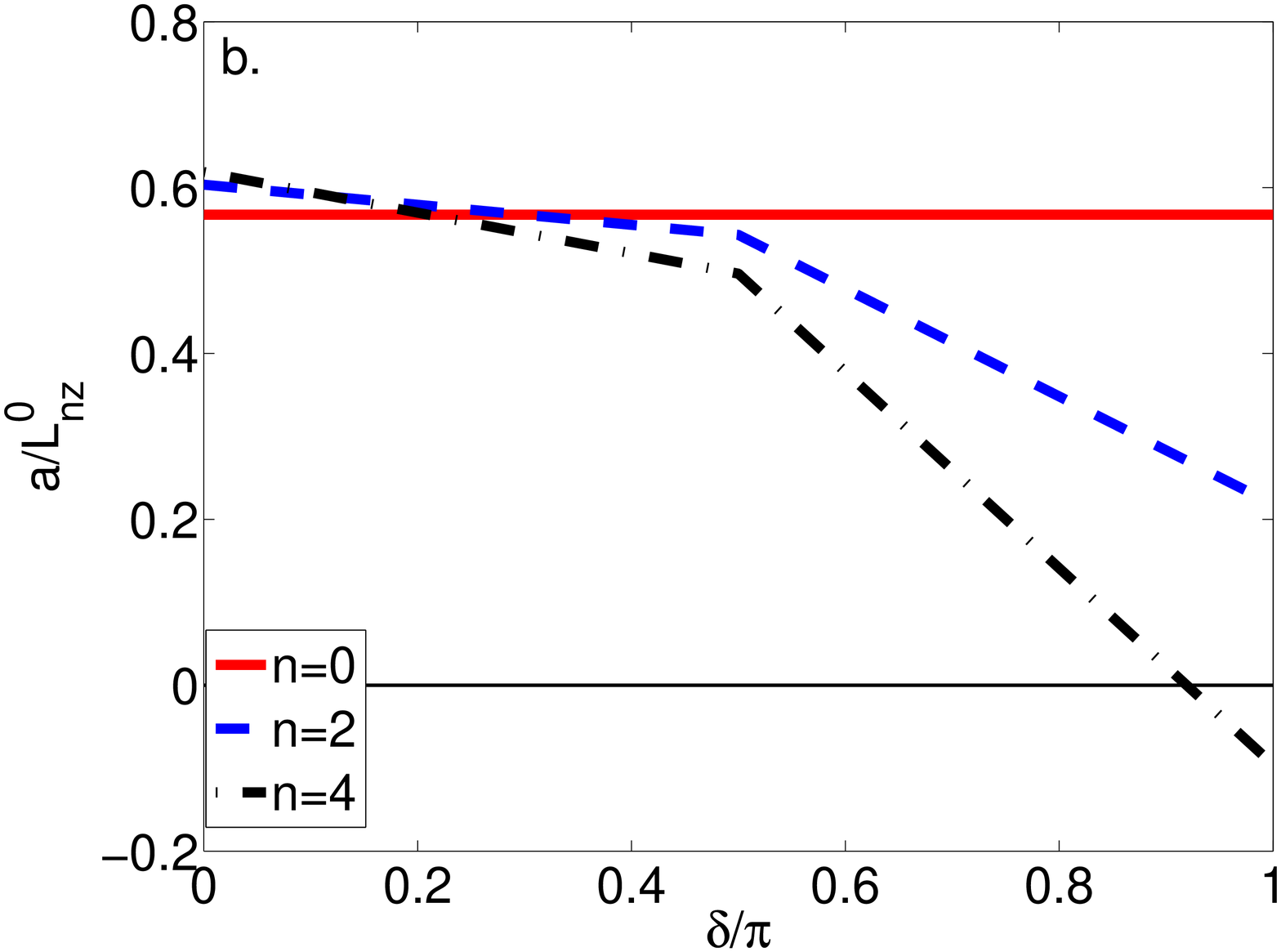}
 \caption{ Peaking factor for nickel as a function of $n$ (a) and
   $\delta$ (b), without taking into account parallel compressibility. (a):
   symmetric impurity density (solid, red), up-down asymmetry (dashed,
   blue), and
   in-out asymmetry (dash-dotted, black). (b): in-out asymmetry -- symmetric
   impurity density (solid, red), $n=2$ (dashed, blue), $n=4$
   (dash-dotted, black). }
\label{nickelwoparacomp}
\end{center}
\end{figure}

It has been shown previously that when the transport is TE-mode
dominated parallel compressibility effects generate an outward
contribution to impurity anomalous flux which can, under certain
plasma conditions, cancel out the inward contributions, leading to
zero or even negative impurity peaking factor
\cite{angionipop07,ap}. In this subsection, we examine this effect by
neglecting the parallel compressibility terms, i.e. terms proportional
to $\delta_p$ in Eqs.~(\ref{pfn}) and (\ref{pfT}).  Note that in this
case only the absolute value of the potential enters in the expressions,
while in the case with parallel compressibility both the imaginary and
real parts and their derivatives are important.  Figure
\ref{phi2woparcomp} illustrates the square of the electrostatic potential
$|\phi|^2$ for different values of $a/L_{Ti}$ and
$a/L_{ne}$.  
We note that the absolute value of the potential is not significantly
different in the various cases, so therefore we do not expect dramatic
dependence on the temperature and density gradients.  Figure
\ref{nickelwoparacomp} shows the peaking factor for various $\delta$
and $n$. The effect of poloidal asymmetry in this limit is in
agreement with our previous work in Ref.~\cite{poloidal} and also, is
similar to that in the previous sections where the parallel
compressibility effects are considered. As seen in
Fig.~\ref{nickelwoparacomp}a, in the absence of parallel
compressibility effects an in-out asymmetry will lead to a negative
peaking factor (outward impurity flux).  An increase of the poloidal
in-out asymmetry will increase the outward flux of impurities as shown
in Fig.~\ref{nickelwoparacomp}b. Note that in the case with parallel
compressibility, the sign change in the peaking factor occurs for
broader range of $\delta$ and for lower asymmetry strength (compare
Figs.~\ref{nickel} and \ref{nickelwoparacomp}).  Also an up-down
asymmetry can lead to a slight reduction of the impurity peaking
factor. This is different from the case where the parallel
compressibility was taken into account, see Fig.~\ref{nickel}a.

\begin{figure}[htbp]
  \begin{center}
\includegraphics[width=0.45\textwidth]{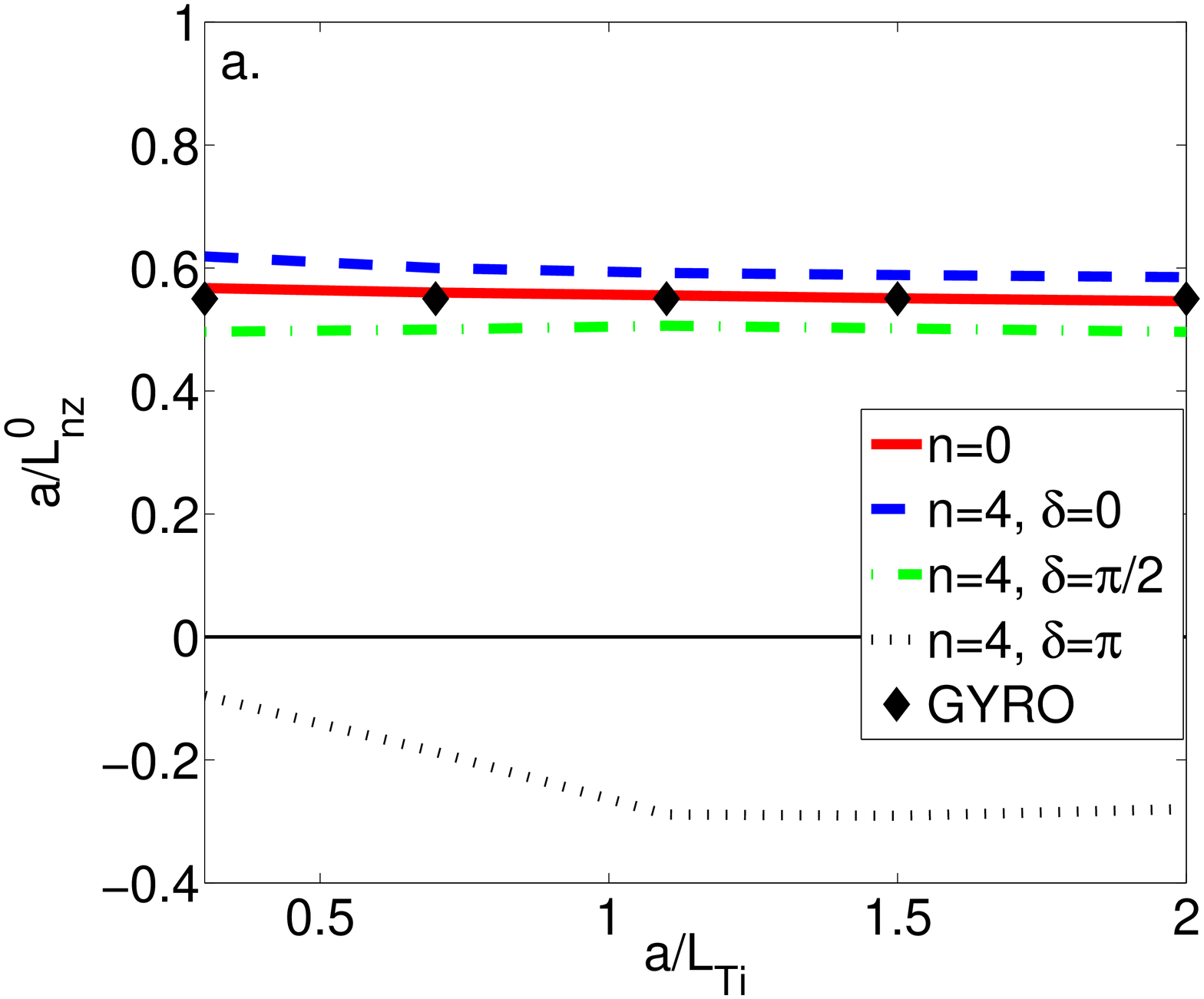}
\includegraphics[width=0.45\textwidth]{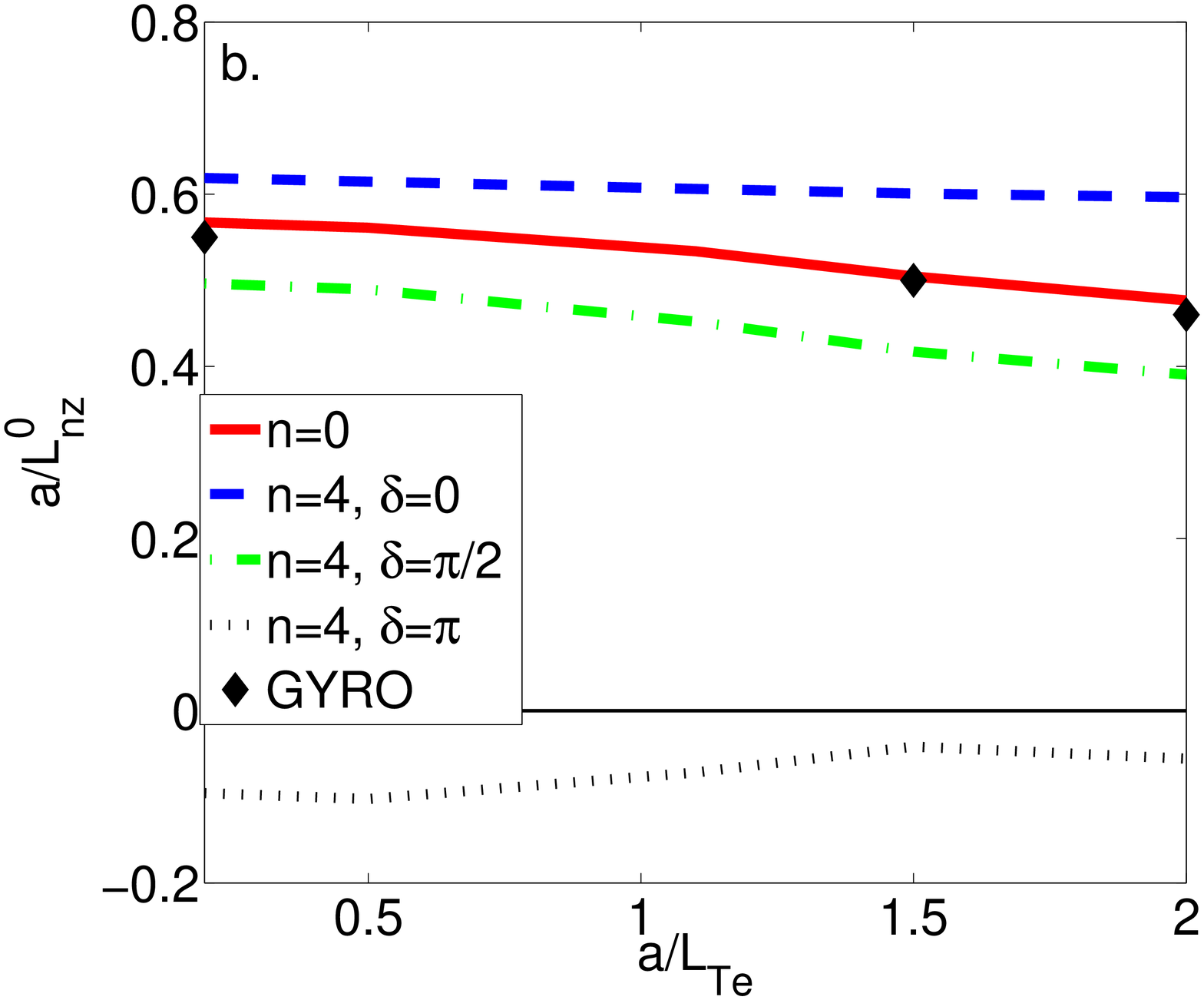}\\
\includegraphics[width=0.45\textwidth]{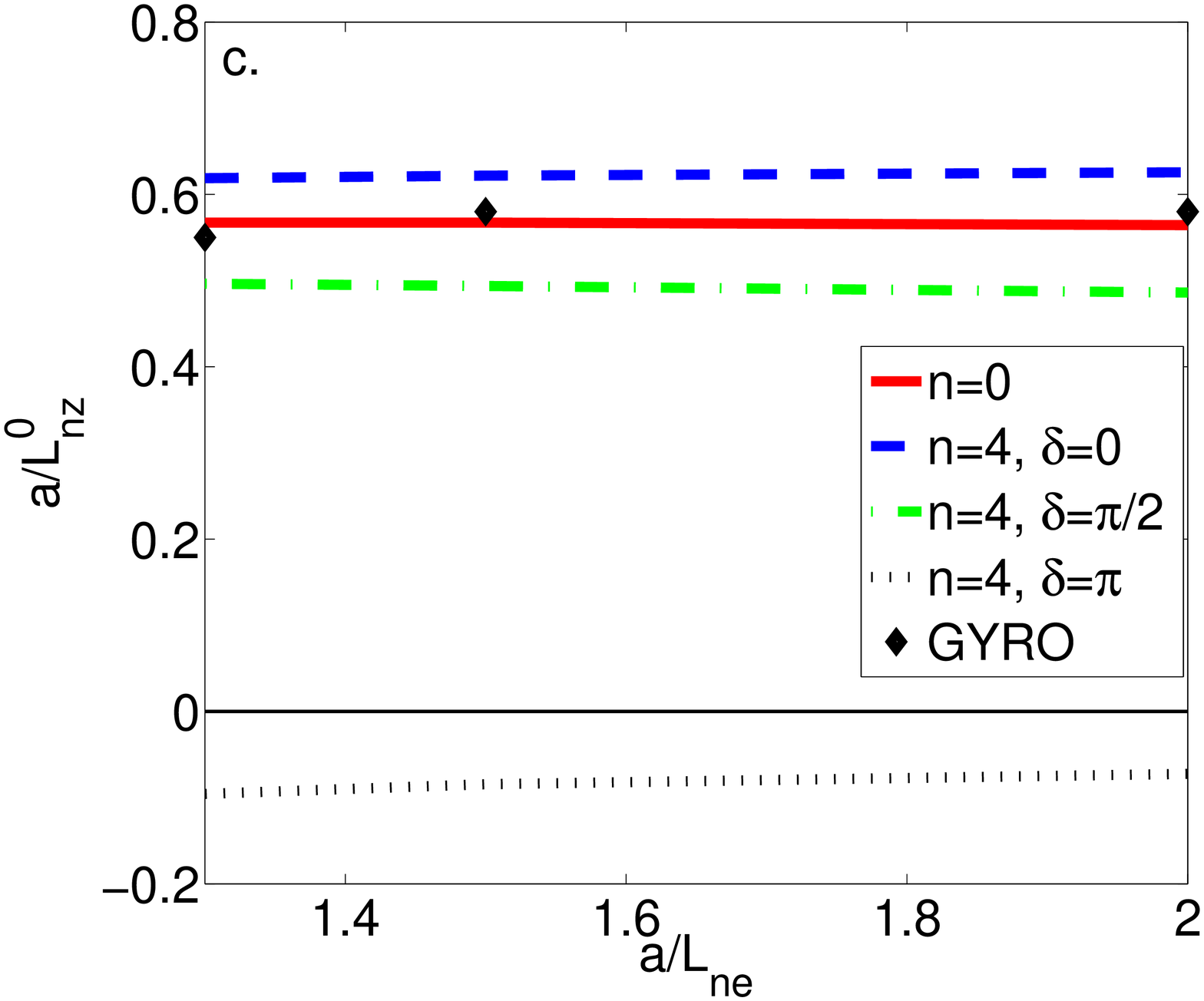}  
\includegraphics[width=0.45\textwidth]{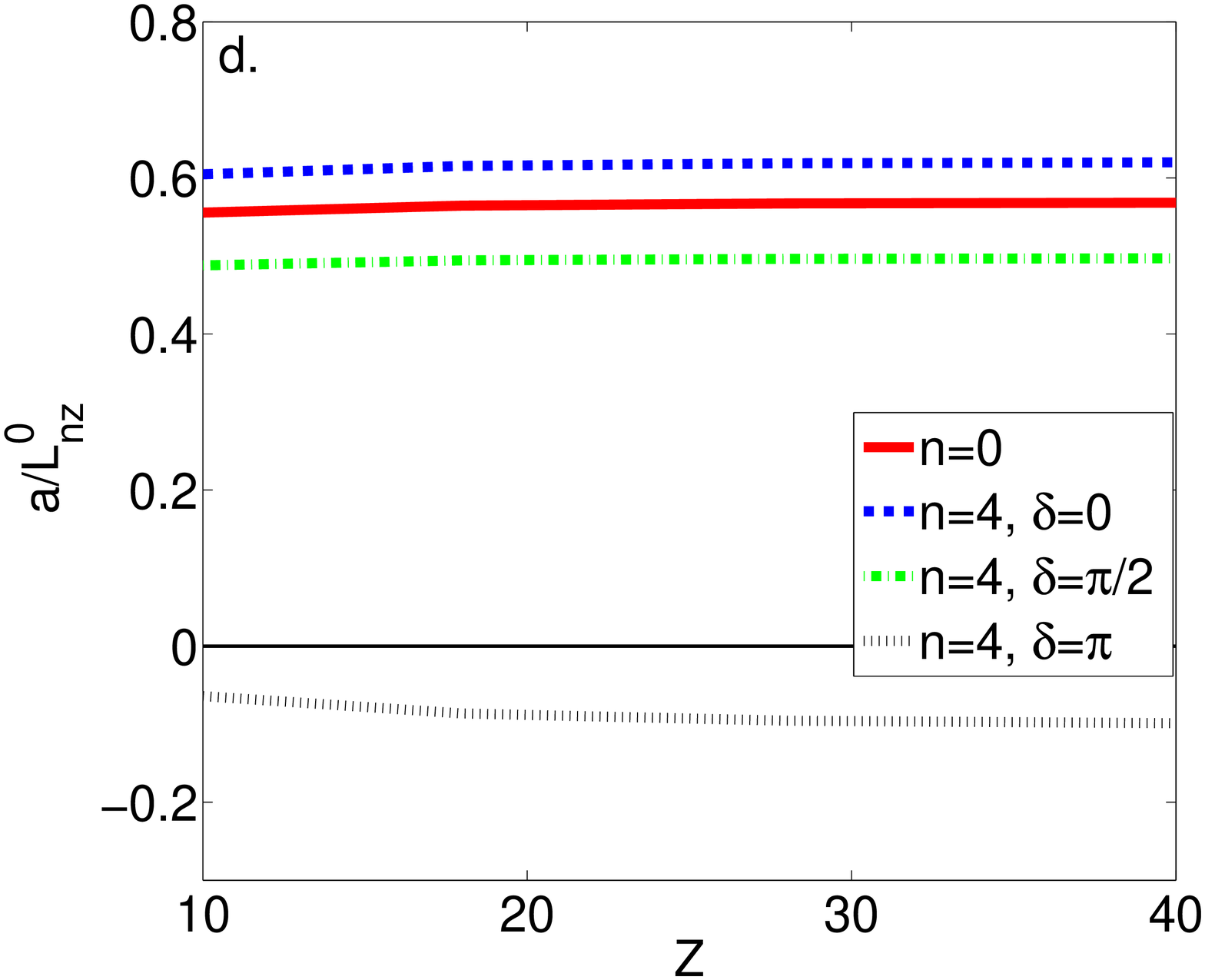} 
    \caption{ Peaking factor for nickel as a function of ion
      temperature gradient (a), electron temperature gradient (b)
      electron density gradient (c) and impurity charge $Z$ (d) for
      different values of $\delta$, without taking into account
      parallel compressibility. Solid (red) line represents the case of
      poloidally symmetric impurity distribution; in figures (a-c)
      this is compared to \gyro simulations (black diamonds). Out-in
      asymmetry (dashed, blue) up-down asymmetry (dash-dotted, green), in-out
      asymmetry (dotted, black). }
\label{tgradscanwoparcomp}
\end{center}
\end{figure}

Figure \ref{tgradscanwoparcomp} shows the density- and
temperature-gradient and charge number scans for the peaking factor
without parallel compressibility. The diamonds represent the values of
peaking factor obtained by \gyro (without parallel compressibility)
which show agreement with our results in the symmetric limit.  From
Fig.~\ref{tgradscanwoparcomp} it is clear that without parallel
compressibility the effect of a poloidal asymmetry is almost
insensitive to the increase of the gradients of temperatures and
density or the impurity charge. This is mainly due to the fact that
without parallel compressibility only the absolute value of the
potential matters and that is rather similar in the various cases.

\section{Discussion and conclusions}
\label{sec:conclusions}
For tokamak operation the avoidance of central impurity accumulation
is a key issue. Conditions in which the convective impurity flux is
directed outward are particularly interesting.  In order to find such
conditions many experiments have been devoted to explore various
techniques. One way to expel the impurities from the plasma core is to
maintain sawtooth crashes in a controlled way by applying central
ICRH~\cite{puiatti02,nave}. It is observed that this method will
indeed remove the impurities from the plasma core. However, it was
shown that even though sawtooth crashes hamper the accumulation of the
impurities their contribution is less relevant compared to the effect
of the ICRH itself~\cite{puiatti03}. Another technique which has been
successful in removing the impurities from the plasma core and is
routinely used in tokamak experiments such as in the ASDEX-U tokamak
is the application of a very localized central
ECRH~\cite{Grubner}. Simulations with both linear and non-linear
gyrokinetic models have shown that under these conditions the electron
temperature gradient is strongly peaked and therefore, modes
propagating in the electron diamagnetic drift direction are the
dominant instability responsible for the turbulence driven transport
\cite{angioniNF09}. The interaction between the related electrostatic
potential fluctuations and the parallel dynamics of the impurities
leads to an outward convection of impurities. These results are in
agreement with experimental observation in ASDEX-U for the very
central region ($r/a\simeq0.2$)~\cite{angionipop07}. In other tokamak
experiments for example at JET the application of the central RF
heating has also been explored with success. In these experiments it
has been observed that the dominant instabilities are not TE-modes but
rather modes directed in the ion diamagnetic drift direction,
ITG-modes, which theoretically should result in an inward impurity
flux. It has been debated that in the very central region of plasmas
at JET the transport is mostly driven by neoclassical effects and by
applying a strong central heating as the temperature gradient peaks
the neoclassical temperature screening effects become dominant and
result in an outward directed impurity flux. Under some plasma
conditions these effects may be the reason for the observed behavior
\cite{duxppcf} however, usually the observed transport is an order of
magnitude larger than the neoclassical predictions implying that the
impurity transport is turbulence driven \cite{angioniNF09}.

In the present work we assume that the impurity density is in the form
of $n_{z}(\theta)=n_{z0}\mathcal P(\theta)$ where the poloidal
dependence is represented by $\mathcal
P(\theta)=\left[\cos^2{\left(\frac{\theta-\delta}{2}\right)}\right]^n$,
and discuss the impact of such a dependence on the impurity peaking
factor in the core of tokamak plasmas. Various mechanisms may give
rise to a poloidal asymmetry of impurity density: difference in
impurity source location, toroidal rotation or neoclassical
effects. Among these is an in-out poloidal asymmetry observed in
plasmas where RF heating is applied. The mechanism responsible for
this behavior was explained through RF-induced accumulation of
minority ions on the outboard side of the torus giving rise to a
corresponding impurity accumulation on the inboard side as was
discussed in Sec.~2. The strength of the impurity accumulation depends
on the impurity charge (higher for heavier impurities) and the plasma
parameters such as temperature gradient. However, the exact form of
these dependences is not yet known and further analysis is needed in
this area.

We have found that an in-out poloidal asymmetry of the impurity
density ($\delta=\pi$) can lead to an outward impurity flux (negative
peaking factor) in both ITG and TE mode dominated cases, and this
effect becomes stronger as the asymmetry strength ($n$) increases. In
this case the peaking factor is strongly charge dependent and it
becomes more negative for heavier impurities. In TE mode dominated case
an up-down asymmetry can also generate a negative peaking factor. This
can be a contributing reason for the observed flat impurity density
profiles in plasmas with ECRH, and up-down asymmetries have indeed
been observed in EC heated plasmas \cite{condrea}, although the
physical mechanism for these asymmetries is not entirely understood.

The reason for the sign change of the impurity peaking factor in the
presence of a poloidal asymmetry is attributed to the interaction
between the poloidal variation of the related electrostatic potential
and the poloidal dependence of the impurity density. If parallel
compressibility effects are taken into account the imaginary part of
the electrostatic potential is the determining factor in the sign
change of the impurity peaking factor.

In summary, our results suggest that poloidal asymmetries can
significantly alter the  turbulence driven impurity transport and
therefore, have to be taken into account. These asymmetries may have a
significant role in determining the impurity accumulation properties
in plasmas with radio frequency heating. Therefore, there is a strong
need for development of new tools in order to detect poloidal
asymmetries and determine the effect of RF heating in their generation
and the dependence of the asymmetry function to various plasma
parameters.

\section*{Acknowledgments}
The authors would like to thank J Candy for providing the \gyro code.
This work was funded by the European Communities under Association
Contract between EURATOM and {\em Vetenskapsr{\aa}det}. The views and
opinions expressed herein do not necessarily reflect those of the
European Commission.

\end{document}